\documentclass[trackchanges,twocolumn]{aastex701}
\usepackage{array}

\newcolumntype{M}[1]{>{\centering\arraybackslash}m{#1}}
\newcolumntype{N}{@{}m{0pt}@{}}

\usepackage{graphicx}	
\usepackage{amsmath}	
\usepackage{mathrsfs}	
\usepackage{subfig,caption}
\usepackage{subcaption}
\usepackage{wrapfig}
\usepackage{booktabs}
\usepackage{multirow}

\begin{document}

\title{Uncovering New 
Ionized Winds 
and Relativistic Hot Outflows in MCG-6-30-15 with \emph{Chandra} HETG and a Bayesian Framework}

\author[orcid=0000-0003-3496-8928,gname='Erika B.',sname='Hoffman']{E.~B.~Hoffman}
\affiliation{Department of Astronomy, University of Maryland, College Park, 20742, USA}
\affiliation{X-ray Astrophysics Laboratory, NASA Goddard Space Flight Center, Greenbelt, MD 20771, USA}
\affiliation{Center for Research and Exploration in Space Science and Technology, NASA/GSFC (CRESST II), Greenbelt, MD 20771, USA}
\email[show]{ebhoff@umd.edu} 

\author[orcid=0000-0003-4504-2557,gname='Anna.',sname='Ogorza{\l}ek']{A. Ogorza{\l}ek}
\affiliation{Department of Astronomy, University of Maryland, College Park, 20742, USA}
\affiliation{X-ray Astrophysics Laboratory, NASA Goddard Space Flight Center, Greenbelt, MD 20771, USA}
\affiliation{Center for Research and Exploration in Space Science and Technology, NASA/GSFC (CRESST II), Greenbelt, MD 20771, USA}
\email{ogoann@umd.edu} 

\author[orcid=0000-0002-1510-4860,gname='Christopher S.',sname='Reynolds']{C.~S.~Reynolds}
\affiliation{Department of Astronomy, University of Maryland, College Park, 20742, USA}
\affiliation{ Joint Space-Science Institute, College Park, MD 20742-2421, USA}
\email{creynold@umd.edu} 

\author[orcid=0000-0002-5466-3817,gname='Lia.',sname='Corrales']{L. ~Corrales}
\affiliation{Department of Astronomy, University of Michigan, Ann Arbor, MI 48109, US}
\email{liac@umich.edu} 

\begin{abstract}
Active Galactic Nuclei (AGN) feedback plays a key role in galaxy evolution. Highly ionized outflows, detected in X-ray absorption, are a promising candidate for quasar-mode feedback. We present a new Bayesian analysis of 0.65 Ms of \emph{Chandra} High Energy Transmission Gratings (HETG) observations of MCG-6-30-15 from 2000 and 2004. Our approach detects new wind components, improving estimates of outflow demographics and impact. These include a hot ultra-fast outflow (UFO; 0.08c), recently confirmed by XRISM but seen in HETG for the first time, which has the power to potentially influence its host galaxy. 
We also detect the first potentially collisionally ionized absorbers in this source, a physical process typically not considered in previous analyses, but potentially important for wind impact and demographics.
The warm absorber complex is resolved in unprecedented detail, uncovering evidence of a correlation between outflow velocity and ionization, constraining its geometry. We also confirm the presence of dust in the wind, clearly distinguishing it from the cold ISM components, which may impact wind acceleration.  Our work highlights the power of applying new methods to deep, legacy HETG datasets, and the limitations of current atomic databases when considering OIV, V, and VI K-shell transitions. Furthermore, our work establishes a baseline for decade-scale outflow variability for new observations with XRISM and NewAthena in the future.

\end{abstract}




\section{Introduction}

Uncovering the nature of Active Galactic Nuclei (AGN) feedback is crucial to understanding the evolution of galaxies \citep{2012ARA&A..50..455F, 2024Galax..12...17H}.
AGN are highly luminous ($10^{43}$ -- $10^{46}$ erg s$^{-1}$) sources located at the centers of many large galaxies, powered by matter accretion onto a supermassive black hole \citep[SMBH,][]{1984ARA&A..22..471R}. SMBHs likely exist at the center of all galactic bulges \citep{2014ARA&A..52..589H}, and exhibit evidence for a critical relationship with their galactic hosts \citep{2012ARA&A..50..455F}. This evidence takes many forms, some of the most notable being that SMBH mass correlates with various host galaxy properties across cosmic history, suggesting co-evolution \citep{2000ApJ...539L..13G, 2003ApJ...587...25D, 2012ARA&A..50..455F, 2014ARA&A..52..589H}. Furthermore, AGN feedback via heating gas or expelling gas from the interstellar medium and quenching star formation is required in cosmological models to produce the observed properties of massive galaxies \citep{2012RAA....12..917S, 2014ARA&A..52..589H}. 
Broadly, AGN radiation, winds, and jets 
have a strong potential to explain AGN feedback \citep{2012ARA&A..50..455F}. 

In particular, highly ionized winds, wide-angle matter outflows launched near the SMBH accretion disk are likely responsible for AGN feedback in radiative-mode \citep[where $L_{Bol}/L_{Edd} > 1\%$,][]{2012ApJ...753...75C, 2012MNRAS.422L...1T, 2021NatAs...5...13L}. 
Although often spatially unresolvable, these ionized winds are studied through the blueshifted absorption lines they imprint on UV and X-ray continuum spectra \citep[e.g.][]{2011A&A...534A..38D, 2023A&A...680A..44L, 2025Natur.641.1132X}. 
Such studies can directly measure wind ionization, column density, line width, and outflow velocity. 
X-ray wind absorption is common, seen in $\sim$40-60\% of unobscured, radio-quiet AGN sources \citep{1999ASPC..175..341C, 2013MNRAS.430.1102T, 2014MNRAS.441.2613L}. 
Moderate-luminosity AGN ($10^{43}$ -- $10^{45}$ erg $s^{-1} $) 
have the potential to contribute significant feedback to their host galaxy environments \citep{2012ApJ...753...75C, 2013MNRAS.430.1102T}, with estimates of total wind kinetic luminosity $L_{KE}$ in ranges models predict can regulate SMBH and galactic bulge growth \citep[0.5-5\% $L_{KE}$/$L_{bol}$,][]{2010MNRAS.401....7H}.

Among the AGN outflows found in UV and X-ray absorption spectra, the winds exhibit a wide range of properties, suggesting potential sub-types.
Wind components are generally classified into 4-6 main categories: Narrow and Broad Absorption Line Outflows, typically seen in the rest frame UV spectra of quasars with low ionizations (log($\xi$/[erg cm s$^{-1}$]) $<$ 2.5), and are differentiated by their line widths (NALs, $\sigma$ $<$ 500 km s$^{-1}$ \& BALs, $>$2000 km s$^{-1}$; \citealt{1981ARA&A..19...41W}) and velocities (NALs, usually v$_{out}$ $<$ 4000 km s$^{-1}$, BALs, $>$ 5000 km s$^{-1}$). When UV winds exhibit properties between both types, and are classified as mini-BALs \citep{2004ASPC..311..203H}. In X-ray spectra, Warm Absorbers (WAs) and Ultra-Fast Outflows (UFOs) are most commonly seen. WAs have low speeds and ionizations, with some resemblance to NALs, except for their higher column densities (log(N$_H$/[cm$^{-2}$]) $>$ 20). UFOs frequently exhibit mildly relativistic speeds (0.1--0.3c), similar to BAL speeds, but with much higher ionizations (log($\xi$/[erg cm s$^{-1}$]) $>$ 2.5), and a wider range of line widths. Occasionally, some low-ionization parameter UFOs (low-IP UFOs) are detected in X-rays \citep[e.g.][]{2013ApJ...772...66G, 2015ApJ...813L..39L, 2024A&A...687A.179X, 2024ApJS..274....8Y}, which, other than their narrow line widths, bear a close resemblance to BALs. It is thought that these different wind sub-types may have different origins and launching mechanisms, or they are closely related, but observed at different evolutionary stages or spatial scales. UFOs and BALs are often argued to be nearest to the black hole (0.001-10 pc away, UFOs due to their high ionization, BALs due to their short variability timescales \citealt{2013ApJ...777..168F, 2021NatAs...5...13L}) while low-variability BALs have the potential to be at much greater scales than UFOs (up to hundreds of parsecs, \citealt{2013ApJ...777..168F}). Due to their much lower ionizations, BALs are likely shielded from the hard X-ray radiation of the central engine, perhaps by an X-ray wind \citep{1995ApJ...451..498M}. WAs and NALs may be at greater distances than UFOs and BALs, likely related to the Broad Line Region, the torus, the narrow line region, or farther in the galaxy \citep{2013MNRAS.430.1102T, 2021NatAs...5...13L}. 

In the traditional picture of UV and X-ray absorbing winds, the physical process assumed to dominate all the absorbing gases is photoionization, due to the extreme luminosities of AGN, and are most commonly assumed to be well approximated as being in photoionization equilibrium (PIE). 
Collisional ionization has long been suspected to also explain some of the spectra \citep{1999ApJ...512..184N}, however, only recent work have been able to distinguish that some winds may be better modeled in collisionally ionized equilibrium \citep[CIE,][]{2022arXiv220808457O, 2025ApJ...979..101T, 2025arXiv250111562M}.  
Even for some previously detected absorbing components, CIE models can sometimes better represent a collection and/or a wider range of spectral features that were previously modeled as PIE, suggesting different origins, locations, or physical processes. For instance, CIE gas may be shielded from the central engine, shock-heated, and/or part of the hot ISM farther in the galaxy, with a less direct connection to the SMBH. 

Differentiating the spectral data's preference for a model in CIE or PIE for a given wind is possible in many cases. Still, it is non-trivial since it requires the highest spectral resolution data available, accompanied by an often computationally intense, agnostic search of parameter space and statistically-motivated model selection. Simultaneously, this approach is also advantageous in finding all detectable winds in a given source, which has been in some sources found to be $\sim$10 absorbing components (e.g. \citealt{2025A&A...694A.302L}).

The soft X-ray band (0.5-2 keV) exhibits the most absorption features, typically from low-velocity gas and low-IP UFOs, while the hard X-ray band (2-10 keV) contains lines from fast, high-ionization winds and UFOs.
\emph{Chandra X-ray Observatory’s} High Energy Transmission Gratings (HETG) have the highest spectral resolution in the soft X-ray band ($\Delta E /E$ $\sim$1000 at 1 keV), and simultaneous coverage of the hard X-ray band. The recently launched X-ray Imaging and Spectroscopy Mission \citep[XRISM,][]{2022SPIE12181E..1SI} has the highest resolution in the hard X-rays, but no soft-band coverage due to the currently closed gate valve. HETG observations from early in its mission, before the soft X-rays became absorbed by contamination \citep{2005SPIE.5898..313O}, are the best existing soft-X-ray high-resolution data with simultaneous hard-X-ray data. Combined with long exposure times and a large archive, HETG allows us to obtain tight constraints on X-ray wind properties, to better understand their ionization process, total number, variability, and power.

Here, we study the 130 ks and 520 ks exposures from 2000 and 2004, respectively, of \emph{Chandra} HETG observations of the Seyfert I galaxy MCG-6-30-15 (hereafter MCG-6). This is one of the deepest early mission \emph{Chandra} HETG data sets, ideal for conducting an in-depth analysis of X-ray-detected AGN outflows. MCG-6 is a nearby (z = 0.007749, \citealt{1995ApJS..100...69F}), luminous ($L_{bol}$ $>$ $10^{44}$ erg s$^{-1}$, \citealt{2009MNRAS.399.1553V}),
$10^6 M_\odot$ SMBH \citep{Bentz_2016} exhibiting persistent winds \citep{2001ApJ...554L..13L, 2005ApJ...631..733Y, 2010ApJ...708..981H, 2011MNRAS.414.2345C}. 
Furthermore, since the \emph{Chandra} archive contains 2 datasets, and MCG-6 is a small $10^6 M_\odot$ black hole (and therefore smaller emitting regions and disk timescales), we can study variability between the epochs of 2000 and 2004.

MCG-6 is also the first AGN with detection of cold (solid or neutral/low ionization) Fe, detected via its L-shell edge in both HETG and XMM-Newton Reflection Grating Spectrometer (RGS) data \citep{2001ApJ...554L..13L, 2003ApJ...596..114S}, and high optical reddening \citep[E(B-V)=0.61-1.09,][]{1997MNRAS.291..403R}. The cold iron in the rest frame of MCG-6 has an unusually high absorbing column, an order of magnitude higher than from the ISM in our Galaxy, while lacking any accompanying detection of neutral O, Mg, or Si edges in previous work \citep{2003ApJ...596..114S}, making the nature of this Fe uncertain. 

Dust has been suggested as a potential driving mechanism for warm absorbing winds, since it can significantly improve the coupling to radiation \citep{2008MNRAS.385L..43F, 2018MNRAS.476..512I}. Therefore, differentiating whether these MCG-6 Fe L features originate from dust or gas is an important prospect. The potential for Fe dust to be embedded in a warm absorbing wind \citep{2001ApJ...554L..13L} instead of the ISM of MCG-6 may be one of the most direct connections between dust and X-ray winds, supporting the mechanism for radiative driving of warm absorbers.

In this work, we apply a Bayesian analysis framework to analyze high-resolution spectroscopic observations, since it is particularly equipped for statistically motivating selection between nested and non-nested models which are equally physically plausible \citep{2001ApJ...548..224V}. Furthermore, it is effective at computing error bars and parameter degeneracies for complex data and models.
We search for all absorption components detectable in the data with large ionized gas models. We are the first to consider both collisional and photoionization processes in this source, uncovering several components that were previously unrecognized.

In Section \ref{data}, we discuss the observations and data reduction. In Section \ref{meth}, we outline our Bayesian framework and how we apply it to spectra models of both the continuum emission and absorption. In Section \ref{res} we present results for broadband components, Galactic absorption models, and MCG-6 absorption. In Section \ref{diss}, we discuss our new results in comparison to previous works, estimate the total wind power from our detections, and the limitations of our approach. In Section \ref{conc}, we summarize our conclusions.

All uncertainties are 68\% credible intervals and 90\% upper limits, unless otherwise noted.


\begin{table}
	\centering
	\caption{Good Time Intervals (GTI) for each observational ID.
 }
	\label{tab:gtis}
	\begin{tabular}{lccr} 
		\hline
		OBSID & DATE & GTI [ks] & Total [ks] \\
		\hline
		433\_001 & April 5, 2000 & 65.8 &  \\
		433\_004 & August 21, 2000 & 65.4 &  \\
		 &  & &  131.2\\
		4759 & March 19, 2004 & 161.1 &  \\
		4760 & March 22, 2004 & 165.2 &  \\
		4761 & March 24, 2004 & 158.8 &  \\
		4762 & March 26, 2004 & 38.2 &  \\
		 &  & & 523.3 \\
		 \hline
		 &  & & 654.5 \\
		\hline
	\end{tabular}
\end{table}

\section{Data}\label{data}

\subsection{Observations}
\begin{figure}
    \centering
	\includegraphics[width=\columnwidth]{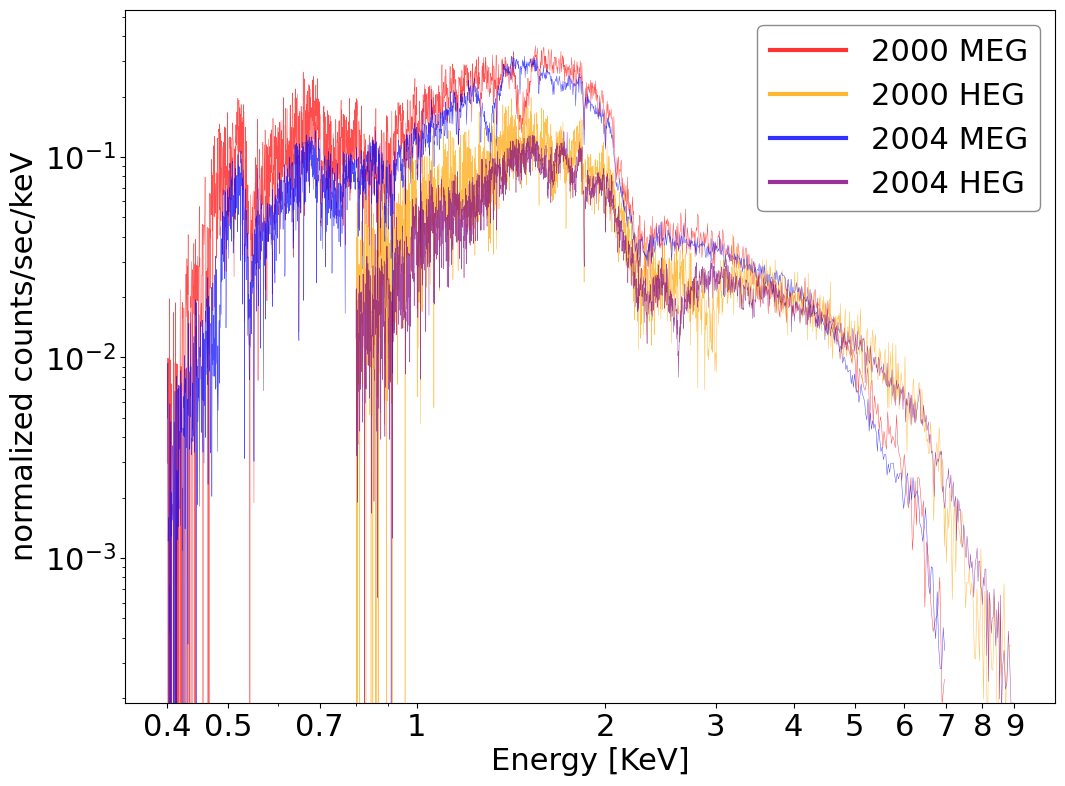}
    \caption{The normalized counts/sec/keV for each \emph{Chandra} High Energy Transmission Grating (HETG) observation, from 2000 and 2004, before the soft energies were significantly blocked by a contaminant \citep{2005SPIE.5898..313O}, for each of the two HETG gratings. The Medium Energy Grating (MEG) has a higher effective area in the soft band down to 0.4 keV, but no response above 7 keV, while the High Energy Grating (HEG) has a range of 0.8-9 keV.}
    \label{fig:norm_counts}
\end{figure}

We use archival \emph{Chandra} HETG observations of MCG-6-30-15, which includes data from both the High Energy Grating (HEG, 0.8–9 keV) and the Medium Energy Grating (MEG, 0.4–7 keV). This includes two observations in 2000 and four observations in 2004, contained in the Chandra Data Archive (CDC)~\dataset[DOI: 10.25574/cdc.535]{https://doi.org/10.25574/cdc.535}. These datasets are summarized in Table \ref{tab:gtis}, and their normalized counts are shown in Figure \ref{fig:norm_counts}. 

\subsection{Data Reduction}

We use the \texttt{CIAO v4.14} software package to reduce the data \citep{2006SPIE.6270E..1VF}. We followed the standard procedures for reducing HETG/ACIS-S Grating Spectra \footnote{https://cxc.cfa.harvard.edu/ciao/threads/spectra\_hetgacis/} except for reducing the spectral extraction width by a factor of 2 and removing flares via \texttt{deflare}. The reduced extraction width allows us to extract the overlapping grating data to higher energies than the standard reduction, resulting in 0.4--7 keV and 0.8--9 keV for MEG and HEG, respectively. 
We removed periods of high background, and the resulting Good Time Intervals (GTIs) are in Table 1.

We grouped each spectrum such that each bin had at least one count. We did not co-add the positive and negative orders nor individual observations, and instead, we fit all spectra simultaneously within each epoch. Adding high-resolution, high signal-to-noise absorption spectra may result in artificial features due to systematics in wavelength calibration \citep{2011A&A...534A..37K}.

\section{Methods} \label{meth}
In this section, we describe the broadband emission and ionized absorption models tested in our analysis, as well as the Bayesian framework we use to select the best models to describe the data.


\subsection{Broadband Continuum Models}

HETG is not ideal for constraining AGN continuum, since it lacks hard X-ray sensitivity; however, we aim to sufficiently represent the emission for accurate absorption modeling. We test select models that represent hot coronae, warm coronae, relativistic reflection, and neutral reflection. The choice of these components is motivated by a large literature on this object using high signal-to-noise, medium resolution X-ray spectra by ASCA, XMM-Newton, and Nustar \citep{1995Natur.375..659T, 2014ApJ...787...83M}, as well as the high spectra-resolution microcalorimeter on XRISM \citep{2025ApJ...995..200B, 2026ApJ..1003..103W}. Unless otherwise noted, all models are part of the XSPEC spectral fitting package, ver. 12.12.1 \citep{1996ASPC..101...17A}. 

We start with a baseline model of hot coronal emission via \emph{powerlaw}. We account for Galactic absorption in the neutral interstellar medium (ISM) with \emph{TBabs} \citep{2000ApJ...542..914W}. We allow for a velocity shift via \emph{zmshift} to take into account the high spectral resolution, potential for moving ISM gas, and $\sim$2 eV uncertainty in atomic data \citep{2024ApJ...965..172C}. We also include a \emph{constant} factor between the two gratings on HETG to account for calibration uncertainties. 

To account for the soft excess emission ($<$2 keV), we consider the phenomenological disk emission with \emph{diskbb} \citep{1986ApJ...308..635M}, and the thermally comptonized continuum \emph{nthComp} \citep{1996MNRAS.283..193Z, 1999MNRAS.309..561Z}. To represent the warm coronae, we fix the \emph{nthComp} seed photon temperature (low-energy rollover, below the HETG band) to $kT_{bb} = 0.003$ keV,
and we limit electron temperature \emph{$kT_e$} $\lesssim$ 0.3 keV to ensure it models the soft-excess \citep{2012MNRAS.420.1848D} and not the hot corona. Similarly, for the \emph{diskbb} component, we restrict the inner disk radius temperature $T_{in} \lesssim 0.7$ keV.

We examine all accretion disk relativistic reflection models within the \emph{relxill} model family \citep{2022MNRAS.514.3965D}, including: the standard model with a high-energy cutoff: \emph{relxill}, a hot thermally comptonized continuum source model: \emph{relxillCp}, and the equivalent lamppost geometry models of each: \emph{relxillp}, \emph{relxilllpCp} (including iongrad\_type = 0, 1, 2). In the comptonized continuum models, the emissivity profile is fit with a broken powerlaw separated by a free breaking radius \emph{R$_{B}$} parameter, with indices $Index1$ and $Index2$. To test for neutral reflection, we use \emph{xillver} and \emph{xillverCp}. Since our spectra do not include energies beyond 9 keV, the hot coronae high-energy power-law cutoff parameters are kept fixed, \emph{Ecut} = 300 keV for \emph{relxill} and \emph{xillver}, and similarly \emph{$kT_e$} = 60 keV for \emph{relxillCp} and \emph{xillverCp} with the comptonized continuum.  We fix the inner boundary of the disk \emph{R$_{in}$} at the Innermost Stable Circular Orbit (ISCO) and the outer boundary at \emph{R$_{out}$} = 400 R$_g$. The redshift is fixed at the redshift of MCG-6. 
For the lamppost models, we assume the coronal velocity relative to the black hole and disk \emph{beta} is zero. 

\subsection{Ionized Absorption Models}
We directly model the absorption due to highly ionized gas obscuring the AGN continuum using transmission curves simulated with the \texttt{Cloudy} spectral synthesis code version c25.00 \citep{2017RMxAA..53..385F, 2025RMxAA..61c.120G}. We consider absorbers in collisional and photoionized equilibrium
and generate large grids of models for physically motivated ranges of gas parameters that produce spectral features in the X-ray band. We use uniform, noninformative priors in our Bayesian analysis. 
We calculate both CIE and PIE models using consistent integer-indexed fine-structure energy levels up to 100 for Fe and 50 for all other elements, which allows for sufficient accuracy while still being computationally efficient. This is especially important for modeling Fe M-shell absorption (near log($\xi$/[erg cm $s^{-1}$]) $\sim$ 3) in PIE, which is not included in the base \texttt{Cloudy} model by default.

\begin{table*}
    \centering
    \caption{Ranges and sampling for ionized gas parameters of model grids calculated with \texttt{Cloudy} c25.00 \citep{2017RMxAA..53..385F}.}
    \label{tab:params1}
    \begin{tabular}{lcccc|lcccc}
        \multicolumn{5}{c}{Photoionized Gas Parameters} & \multicolumn{5}{c}{Collisionally Ionized Gas Parameters} \\
        \cmidrule(lr){1-5} \cmidrule(lr){6-10} 
        Parameter & Unit & Min & Max & Spacing & Parameter & Unit & Min & Max & Spacing \\
        \cmidrule(lr){1-5} \cmidrule(lr){6-10} 
        log Ionization Param. $\xi$  & ergs cm $s^{-1}$ & $-2$ & $6$ & $0.15$ 
            & log Temperature $T$  & K & $4.2$ & $9$ & $0.15$ \\
        log Column Density $N_H$  & cm$^{-2}$ & $19.2$ & $24$ & $0.25$
            & log Column Density $N_H$ & cm$^{-2}$ & $19$ & $24$ & $0.25$ \\
        log Line Width $\sigma$ & km s$^{-1}$ & $1.5$ & $4.5$ & $0.25$
            & log Line Width $\sigma$ & km s$^{-1}$ & $1.5$ & $4.5$ & $0.25$ \\
            \cmidrule(lr){1-5} \cmidrule(lr){6-10} 
    \end{tabular}
\end{table*}

Absorption features in high spectral-resolution data can be sensitive to small, non-linear changes in parameter values, especially ionization parameter and temperature,
so to minimize bias in the linear interpolation, we finely sample each parameter. For the most abundant elements (Fe, O, Ne), our sampling of ionization parameter for photoionized models, or gas temperature for collisionally ionized models, biases the parameter value by no more than 10\%, and typically less than 5\%.

\begin{figure}
    \centering
	\includegraphics[width=\columnwidth]{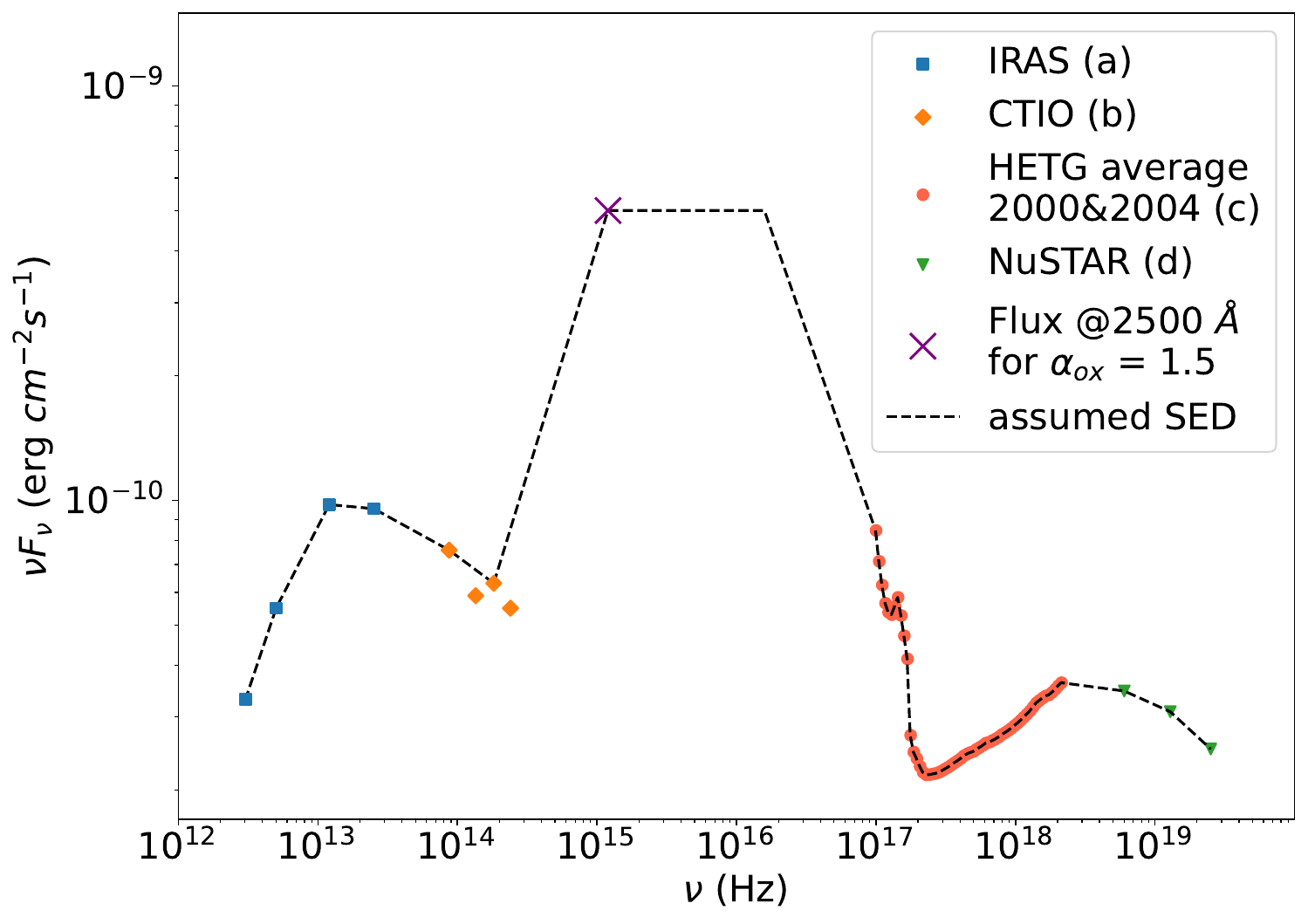}
    \caption{
    The dashed black line shows the assumed broadband SED to calculate \texttt{Cloudy} photoionization models, overlaid with various data sources, including: (a) Infrared Astronomy Satellite (1983, HEASARC), (b) Cerro Tololo Inter-American Observatory \citep{1983ApJS...52..341M}, 
    (c) Chandra HETG (this work) with two datasets' fluxes' averaged to generate one SED for both (d) XMM-Newton and NuSTAR \citep{2014ApJ...787...83M},
    (e) The flux at 2500\textup{~\AA} based on the average HETG flux at 2 keV, assuming $\alpha_{ox}$ = 1.5 \citep{2018A&A...619A..95C}.
    }
    \label{fig:SED}
\end{figure}

\subsubsection{SED for Photoionized Models}\label{sec:sed}
To model photoionized gas, we construct an incident broadband spectral energy distribution (SED) from infrared to hard X-rays, shown in Figure \ref{fig:SED}.
For the infrared portion of the SED, we use fluxes from the literature, including from the Infrared Astronomy Satellite (IRAS) from the Faint Source Catalog (version 2.0) \citep{1990IRASF.C......0M} and the NASA Infrared Telescope
Facility \citep{1987ApJ...315...74W}. Due to the high dust extinction of MCG-6 \citep{1997MNRAS.291..403R}, we cannot directly observe the optical/UV flux. We estimate this portion of the spectrum using an average $\alpha_{ox} \sim$ 1.5 found in studies of large samples of quasars \citep[e.g.][]{2010A&A...512A..34L, 2018A&A...619A..95C, 2023MNRAS.520.2781K}, relating the monochromatic flux at 2500\textup{~\AA} and 2 keV. For the hard X-rays, we use NuSTAR data points \citep{2014ApJ...787...83M}. Although these data are not contemporaneous, it is still correct within an order of magnitude, and it contributes little to the ionization balance compared to the far UV and X-ray bands. 

We use the 2000 and 2004 Chandra HETG data from this work to determine the unabsorbed flux in the 0.4-9 keV range. 
We perform initial fits with a preliminary SED to estimate the impact of wind absorption, and in a second step, we construct the intrinsic SED that is unabsorbed by the outflows.

\subsubsection{Photoionized Equilibrium (PIE) Models}
For photoionized absorbers, we vary grid parameters for column density, line width, and ionization parameter. 
For ionization parameter $\xi$, we consider $10^{-2}-10^6$ ergs cm s$^{-1}$, spacing $\Delta$log$\xi = 0.15$, column density $N_H$ from $10^{19}-10^{24}$ cm$^{-2}$, spacing $\Delta$log$N_H = 0.25$, line width $\sigma$
ranging from $10^{1.5}-10^{4.5}$ cm s$^{-1}$ ($\sim$30 km s$^{-1} -$ 0.1c)
spacing with $\Delta$log$\sigma$ = 0.25 (see Table \ref{tab:params1}). We sample to $\sim$30 km s$^{-1}$ to measure upper limits on absorbers that may be narrower than the instrument's resolution.

\subsubsection{Collisionally Ionized Equilibrium (CIE) Models}
To model collisionally ionized gas absorption, we vary the gas temperature from $10^{4.2} - 10^9$K, in addition to line width and column density, sampled the same as in PIE. We sample temperatures with spacing every log$T = 0.15$ (see Table \ref{tab:params1}).

\begin{figure*}
    \centering
    \includegraphics[width=.75\textwidth]{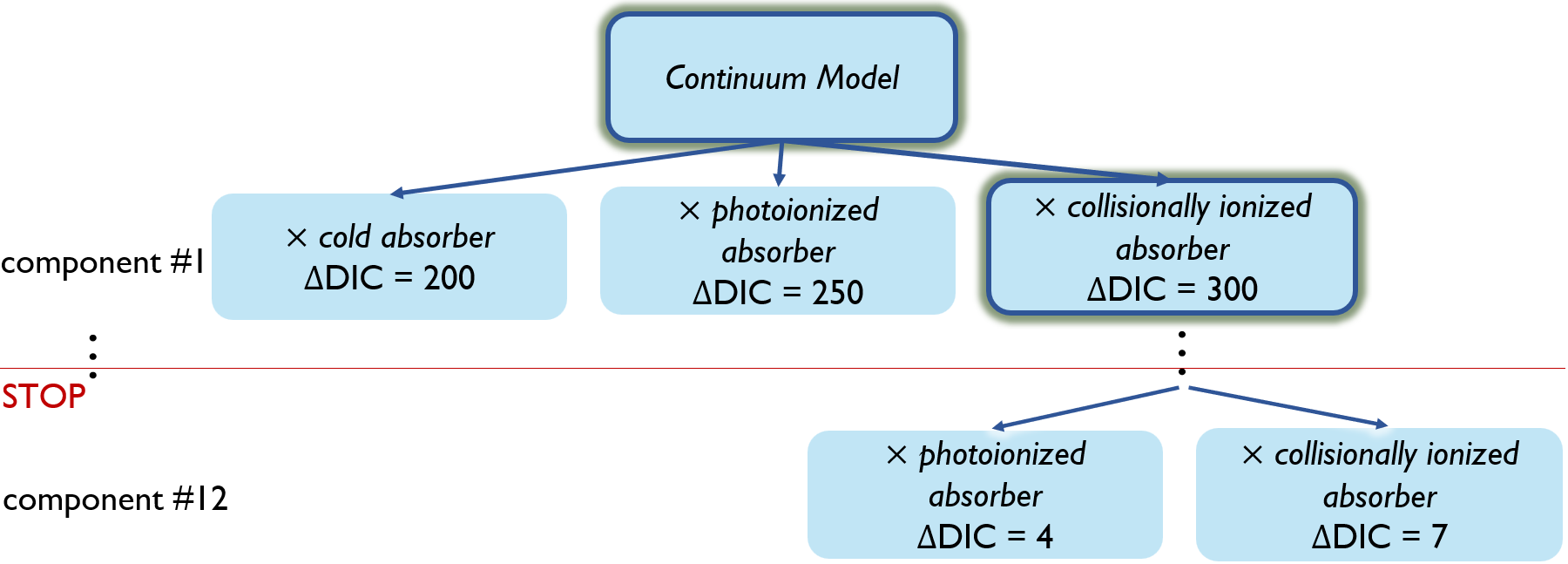}
    \caption{Schematic of an example model selection procedure using Deviance Information Criterion (DIC). The outlined box indicates the largest change in DIC values above the minimum threshold of $\Delta$DIC $>$ 10, and with a relative difference of at least $\Delta$DIC $>$ 10. If both collisional and photoionized components can change the DIC by a similar amount, we report results for both and state that the preference is ambiguous. Finally, we illustrate the clear stopping point in adding more model components when the $\Delta$DIC $<$ 10.
    }
    \label{fig:dic}
\end{figure*}

\subsubsection{Neutral Gas and Dust models}
In addition to modeling Milky Way absorption, we must account for cold material in the target galaxy as well.
MCG-6 is a highly reddened source \citep[E(B-V)=0.61-1.09,][]{1997MNRAS.291..403R}, which implies N$_H$ $\sim$ 4--7 $\times 10^{21}$ cm$^{-2}$ \citep{Cox2002}, and likely requires a model of the cold atomic and/or dusty material along our line of sight. In the X-ray band, this means we may observe the Fe L edge \citep{2001ApJ...554L..13L, 2003ApJ...596..114S}, as well as neutral oxygen gas \citep{2010ApJ...708..981H} and/or dust.
To model the cold Fe and O in MCG-6, we use \emph{tbvarabs}, 
which assumes the cross-sections of atomic O I and solid metallic Fe, which was originally intended to model the fine structure of O K and Fe L shell absorption by the local ISM \citep{2000ApJ...542..914W}. The \emph{tbvarabs} employs photoelectric absorption continuum models from \citet{1995A&AS..109..125V}.

We note, however, that although not tested here since there are no publicly available models for them, neutral iron gas, Fe II and Fe III, and especially iron in different molecules, can yield similar shapes to the assumed metallic iron features and potentially represent this data \citep{2021ApJ...908...52S}.

\subsection{Spectral Fitting and Model Selection}\label{sec:mod}

\begin{table} 
	\centering
	\caption{Jeffery's scale, the interpretation of the $\Delta$DIC, as a measure of significance from comparing model likelihoods ($L_1$ vs $L_2$) on a logarithmic scale \citep{jeffreys1961theory, doi:10.1080/01621459.1995.10476572}. In this work, we conservatively require a `very strong' $\Delta$DIC $>$ 10 to select a model with preference over another, and a minimum change when adding any model to avoid overfitting. On a linear scale, our Bayesian statistic cutoff can be interpreted as: \emph{The data must be at least 150 times more likely under Model 1 than Model 2.}}
	\label{tab:jeff}
	\begin{tabular}{cccr} 
		\hline
		  $\Delta$DIC & & Evidence against a model\\
        =-2$ln(L_1/L_2)$& $L_1/L_2$ &  with higher DIC\\
		\hline
		0--2 &1--3 & Weak  \\
		2--6 & 3--20 &Positive  \\
		6--10 & 20--150 & Strong \\
		$>$10 &$>$150 & Very Strong  \\
		\hline
	\end{tabular}
\end{table}

Our approach follows that of \citet{2022arXiv220808457O}, and extends it to include cold absorbers in addition to ionized ones. For each possible addition of a spectral model component, we determine which competing models are statistically preferred by the data, if at all, in every hierarchical step (see Figure \ref{fig:dic}). For each given model combination we examine, we use the steepest descent algorithm within the \texttt{pyXSPEC} spectral fitting package, ver. 12.12.1 \citep{1996ASPC..101...17A}, with a blind random search, to find the maximum likelihood (best fit) by minimizing the Cash-statistic \citep{1979ApJ...228..939C}.
Secondly, we use a Goodman-Weare Monte Carlo Markov Chain (MCMC) algorithm within XSPEC (\texttt{emcee}) \citep{2010CAMCS...5...65G, 2013PASP..125..306F} seeded near this maximum and run it to convergence. With MCMC, we can calculate the parameter posteriors and covariance for complex data and models, increase confidence in the position of the global maximum, and statistically compare non-nested models \citep{2001ApJ...548..224V}. 

To do the latter, we compute the Deviance Information Criterion (DIC, \citealt{https://doi.org/10.1111/1467-9868.00353}). The DIC is similar to the Akaike information criterion (AIC, \citealt{Akaike1998}), but uses an effective number of parameters instead of an explicit number, yielding a larger penalty for overfitting, and is trivially calculated from MCMC samples. A statistically preferred model generates a lower DIC value, and the difference between two DIC values is a measure of significance in the preference. This procedure is explained further in Figure \ref{fig:dic}. 

We employ a conservative approach according to the logarithmic ``Jeffery's scale" which requires a $\Delta$DIC $>$ 10 (Table \ref{tab:jeff}, \citealt{jeffreys1961theory, doi:10.1080/01621459.1995.10476572}) to indicate statistical significance. If none of the competing nested models can generate at least $\Delta$DIC $>$ 10, as in the right of Figure \ref{fig:dic}, the final model extracts the most information out of the data without overfitting. Alternatively, if two competing non-nested models each yield $\Delta$DIC values that are within 10 of each other, we consider the preference ambiguous. In short, whichever combination of models produces the incrementally and sufficiently ($\Delta$DIC $>$ 10) lowest DIC value is the most statistically supported to represent the data.




\begin{figure*}
    \includegraphics[width=\textwidth]{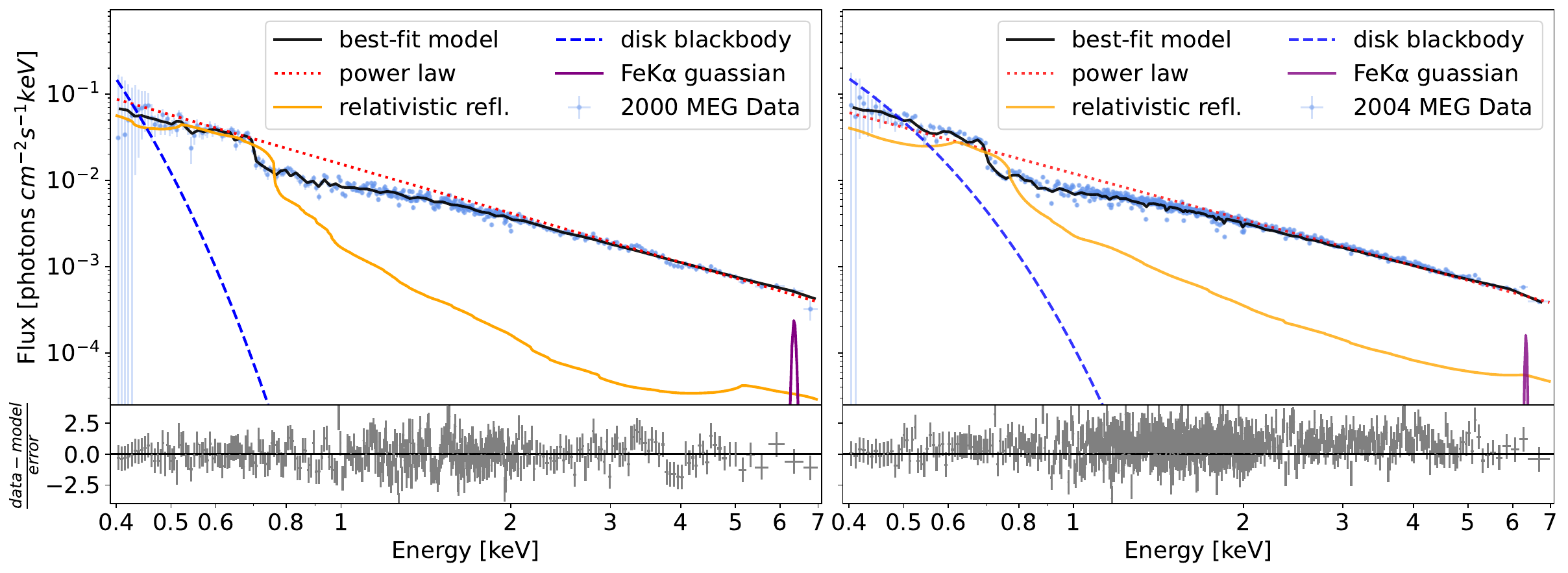}
    \caption{Binned and combined unfolded MEG spectrum of the 2000 (left) and 2004 (right) observations, 0.4-7 keV. The best-fit model (black) is binned more than the data (light blue) to highlight the broadband features. We find the same broadband models best represent both datasets.
    }
    \label{fig:cont}
\end{figure*}

\section{Results}\label{res}

We report the results of our Bayesian analysis for the broadband spectrum in Section \ref{res:cont} and the absorption spectrum in Section \ref{res:abs}. Both the 2000 and 2004 data are best represented by the same broadband X-ray continuum, including relativistic reflection of a Comptonized continuum, and a phenomenological disk blackbody shape to account for the soft excess. 

We detect 6 statistically significant absorbing zones in the 2000 data and 11 in the 2004 data (four to nine ionized in MCG-6, \& one neutral/molecular in MCG-6, one ionized in the Milky Way in each). 
The ionized absorbing gases intrinsic to MCG-6 span a wide range of velocities (inflowing at $\sim$300 km $s^{-1}$ to outflowing at 0.1--0.2c) and ionizations (log($\xi$/[erg cm $s^{-1}$]) $\approx \mbox{-1} -$ 6). Many absorbing gases are equally well or better represented as a gas in collisionally ionized equilibrium (CIE), with measured temperatures ranging from log($T$/[K]) $\approx$4.5--9. We detect a feature in HETG that is likely the hot relativistic outflow seen in XRISM \citep{2025ApJ...995..200B}. Our framework also finds other relativistic components that are likely unphysical and instead may indicate that our calculated models contain incomplete atomic data in the soft X-ray band. Details of all these absorption measurements are described in Section \ref{res:abs}.

\subsection{Broadband Continuum}\label{res:cont}

The broadband spectral shapes for the 2000 and 2004 datasets are shown in Figure \ref{fig:cont}. The illuminating X-ray source for MCG-6 is in a slightly softer flux state in 2000 
with a steeper powerlaw ($\Gamma$ = $1.88 \pm{0.01}$ in 2000 vs $\Gamma$ = 1.77$\pm{0.01}$ in 2004). 

For both the 2000 and 2004 data, the most statistically preferred broadband models, in addition to a neutral galactic-absorbed \emph{powerlaw}, include: relativistic reflection {\emph{relxillCp}, a phenomenological disk blackbody \emph{diskbb}, and a narrow Gaussian line at $\sim$6.4 keV in the rest frame of MCG-6.
\begin{equation}
    \emph{powerlaw + relxillCp + diskbb + gauss@6.4keV} \\
\end{equation}

While we find the preferred broadband models before we begin modeling absorption, \emph{the parameters of each broadband model are left free to vary while modeling absorption}. The final, global best-fit values for these continuum models are reported in Table \ref{tab:cont_best_par}.

\begin{table} 
	\centering
	\caption{Best fit values for broadband continuum parameters with 68\% credible intervals and 90\% upper limits for the global model, including absorption. }
	\begin{tabular}{lccr} 
		\hline
         \hline
        \textbf{Models} & Unit & 2000 & 2004\\
        \& Parameters\\
        \hline
		\hline
         \bf{powerlaw} \\
        \hline
		photon index & - & 1.88 $_{-0.04}^{+0.02}$ & 1.77$_{-0.01}^{+0.005}$  \\
		  \multirow{2}{2em}{normalization} & \scriptsize{$10^{-4} \frac{photons}{keVcm^2s} $} & \multirow{2}{2em}{118$_{-1}^{+1}$} & \multirow{2}{3.5em}{102$_{-3}^{+2}$}  \\
         & \scriptsize{@1 keV}  \\
    	\hline
         \bf{relxillCp} \\
        \hline
        photon index & - & \multicolumn{2}{c}{tied to \emph{powerlaw}}  \\
		disk inclination & $^{\circ}$ & 54 $_{-8}^{+2}$ &  58$_{-6}^{+2}$ \\
		black hole spin & $\frac{cJ}{GM^2}$ &  0.91$_{-0.06}^{+0.03}$ &  0.90$_{-0.1}^{+0.03}$ \\
        break radius & $R_{ISCO}$ & 1.5$_{-0.5}^{+50}$  & 10$_{-5}^{+40}$ \\
        Index1 & - & 4$_{-2}^{+3}$  & 2$_{-1}^{+1}$ \\
        Index2 & - & 5$_{-4}^{+5}$  & 3$_{-1}^{+4}$ \\
        log ionization & erg cm s$^{-1}$ &  3$_{-1}^{+0.2}$ & 2.7$_{-0.2}^{+0.05}$  \\
        log disk density & cm$^{-3}$ &  17.8$_{-0.3}^{+1}$ &  18.8$_{-0.3}^{+1}$  \\
        iron abundance & Solar & $\leq$ 1.3 &  1.5$_{-0.8}^{+0.5}$ \\
        \multirow{2}{4em}{normalization} & \scriptsize{$10^{-5}$$\tfrac{photons}{keV\,cm^2\,s}$} & \multirow{2}{2em}{12$_{-5}^{+2}$} & \multirow{2}{3.5em}{5.4$_{-1}^{+2}$}  \\
        & \scriptsize{@1 keV} & &\\
  		\hline
          \bf{diskbb} \\
        \hline
		$T_{in}$ & eV & $\leq$40 &  76$_{-5}^{+5}$\\
        normalization & \scriptsize{$10^{5}$$({\tfrac{R_{in}}{D_{10kpc}}})^2 cos\theta$} & $\leq$10000 &  2.6$_{-2}^{+2}$ \\
        \vspace{-4mm}\\
        \hline
        \multicolumn{4}{l}{\bf{FeK$\alpha$ Gauss}}\\
		\hline
        LineE (obs) & keV  & 6.34$_{-0.02}^{+0.02}$ & 6.34$_{-0.01}^{+0.01}$    \\
        Sigma & eV & 50$_{-10}^{+10}$ & 22$_{-0.4}^{+1}$    \\
        normalization &  \scriptsize{$10^{-5}$$\tfrac{photons}{keV\,cm^2\,s}$} &  2.8$_{-0.8}^{+0.3}$ & 1.0$_{-0.2}^{+0.2}$    \\
        \vspace{-4mm}\\
		\hline    
         \bf{constant} \\
		\hline
        factor & \% &  97.8$_{-0.6}^{+0.8}$ & 96.5$_{-0.3}^{+0.3}$    \\
        \vspace{-4mm}\\
		\hline    
	\end{tabular}
    \label{tab:cont_best_par}
\end{table}

\subsubsection{Relativistic Reflection}
Due to the limited range ($<$9 keV) and low S/N of HETG in the hard X-rays, the parameters of the reflection models can be biased. Nevertheless, we find that they are still the best description of the analyzed observations.

We find that \emph{relxillCp} (yielding $\Delta DIC$ $>$ 1700 and $>$ 5600 in 2000 and 2004 respectively), is preferred over non-relativistic reflection models (\emph{xillver}, \emph{xillverCp}, \emph{nthcomp}, \emph{diskbb}, with $\Delta DIC$s $>$1100 and $>$5000 in 2000 and 2004 respectively) a with a significantly larger relative improvement in DIC, $>$ 600. The preferred relativistic reflection model is \emph{relxillCp}, compared to \emph{relxill}, a non-Comptonized continuum with a fixed disk density (a greater $\Delta DIC$ $>$ 500 for the 2000 data, and $>$ 1500 for the 2004 data, c.f. Section \ref{sec:mod}).  
Furthermore, we find a preference for the Comptonized continuum model with a broken powerlaw emissivity profile over a lamppost geometry (difference in $\Delta DIC > 30$ for the 2004 data), similar to the analysis of XRISM data \citep{2026ApJ..1003..103W}.

None of the relativistic reflection parameters notably vary between epochs,  and all agree within 1$\sigma$ credible intervals. For both datasets, we find a moderate to high inclination $\sim$54--58$^{\circ}$, a high disk density $\sim$10$^{18-19}$ cm$^{-3}$, an iron abundance near solar, and a high black hole spin $a$  $\sim$0.9 $\frac{cJ}{GM^2}$. 
However, we emphasize the limited coverage of hard X-rays, and our time-averaged approach may bias these values \citep{2026ApJ..1003..103W}. 

\subsubsection{Phenomenological Disk Blackbody}
To account for soft excess not explained by relativistic reflection, we test the addition of non-relativistic models \emph{nthcomp} and \emph{diskbb}. \emph{diskbb} is still able to improve the fit with a $\Delta DIC > 70$ in 2000 and $\Delta DIC > 270$ in 2004 and \emph{nthcomp} produces a $\Delta DIC > 60$ in 2000 and a $\Delta DIC > 260$, compared to just $powerlaw+relxillCp$ alone. We choose \emph{diskbb} therefore because it has a greater change in $\Delta DIC$, but the difference between \emph{nthcomp} and \emph{diskbb} is marginal, only $\sim$10, because \emph{nthcomp} produces a largely similar shape.

\subsubsection{Narrow Fe K$\alpha$ Emission}
We model the weak, narrow Fe K$\alpha$ emission feature with a Gaussian, which is as good a description as global neutral models such as \emph{xillver/xillverCp} \citep{2022MNRAS.514.3965D}, due to the low sensitivity of HETG in this band. Such models also produce unphysical parameters, likely biased due to the treatment of oxygen emission \citep{2025MNRAS.543.2633W}. See Table \ref{tab:cont_best_par} for best-fit values.

\subsubsection{Instrumental Correction Factor}
To account for calibration uncertainties between the two HETG Gratings, we find the multiplicative \emph{constant} factor of 0.975$_{-0.004}^{+0.008}$ in 2000, and 0.965$_{-0.003}^{+0.003}$ should be applied to MEG to best match HEG.




\subsection{Absorption by Winds, ISM and CGM}\label{res:abs}
Observed absorption features can be local to the Milky Way or intrinsic to MCG-6 and come from different gas phases: neutral, CIE, and PIE components. Our Bayesian framework identifies each phase agnostically by order of significance in the data, and we use physical arguments to deduce their galactic or non-galactic origin. We also keep all measured parameters free throughout the entire modeling process to accurately calculate statistical uncertainty.

We first describe the absorption of MCG-6, which exhibits a wide range of physically distinct absorbing zones, comprising 10 zones in the 2004 data and 5 zones in the 2000 data.  These zones include warm absorbers (Section \ref{res:wa}),  up to two equally well modeled in CIE, a fast outflow  potentially in CIE (Section \ref{res:fast}, and a typical hot ultra-fast outflow (UFO).
We briefly mention the likely unphysical cool/low IP components 
fit with relativistic speeds, in Section \ref{sec:pot}.

We also detect in both datasets a Fe L edge, which likely comes from both the MCG-6 ISM and a wind, discussed in Section \ref{res:dust}. Finally, we report the neutral and ionized absorption intrinsic to the Milky Way in Section \ref{res:NMW}. 

\begin{figure}
    \centering
	\includegraphics[width=250pt]{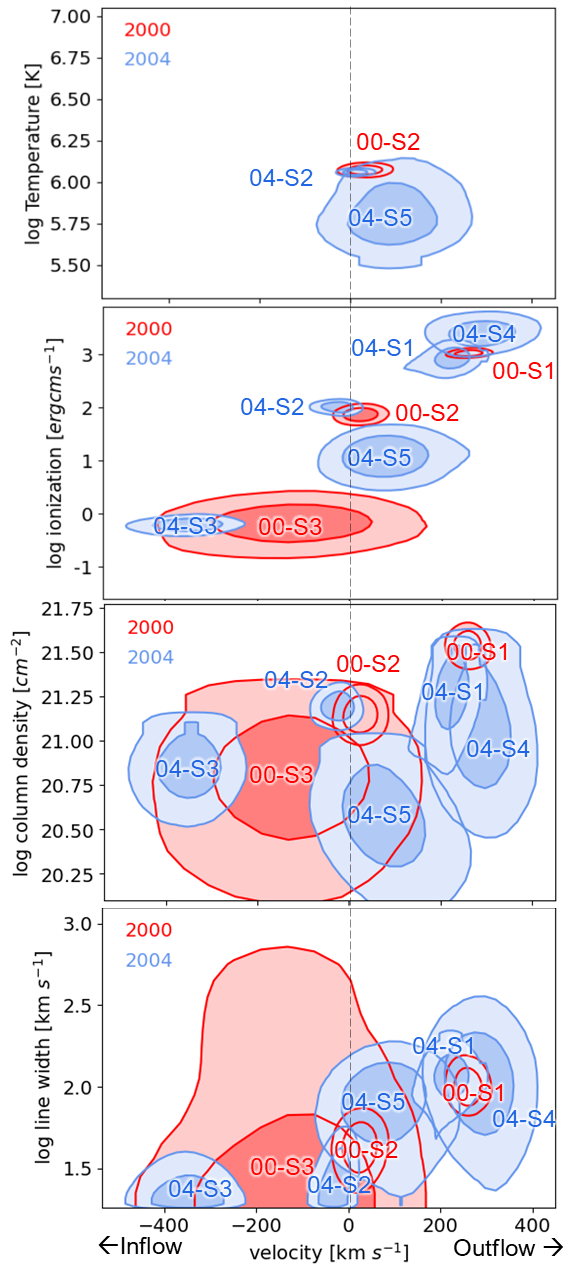}
    \caption{2D posterior probabilities of slow inflow (negative) and outflow (positive) wind components. A dashed line indicates MCG-6's rest frame. Temperature (top) for collisionally ionized (CIE) winds, ionization parameter for photoionized (PIE) winds, hydrogen column density, and line width (bottom) are on the y-axes. The 2000 results are represented in dark red, \& the 2004 in light blue. Contours indicate 1 \& 2$\sigma$ error. The stable WAs show a possible trend between ionization and outflow velocity and are seen as more distinct gases in the deeper 2004 data.  All WAs are  well represented in PIE, but two may also be in CIE.}
    \label{fig:colden}
\end{figure}

\subsubsection{Low-Velocity Inflowing \& Outflowing Warm Absorbers (00-S1--S3, 04-S1--S5)}\label{res:wa}
The second most statistically significant warm absorber in both the 2000 and 2004 datasets is equally well represented in  PIE and CIE, denoted as `00-S2' and `04-S2' in Figure \ref{fig:colden}. This absorber is stable between observations, with a velocity consistent with the rest frame of MCG-6, a column density of log($N_H$/cm$^{-2}$) $\sim 21.2$, and a narrow line width of $\sigma <$ 50 km s$^{-1}$. Assuming it is an absorber in CIE, it would have a temperature log($T/[K]) \sim 6$, or assuming it is an absorber in PIE, it would have an ionization of log($\xi$/[erg cm s$^{-1}$]) $\sim 2$. This component accounts for the majority of the absorption of the oxygen K edge in both datasets, and is likely the main absorber seen in earlier MCG-6 observations \citep[e.g.][]{1997MNRAS.291..403R}.

The other component in 2004, with a velocity consistent within 1$\sigma$ of the rest frame of MCG-6, 04-S3, is a similar narrow-line warm absorber that can also be equally well modeled in CIE or PIE. 
It exhibits a lower temperature than 04-S1, at either log($\xi$/[erg cm s$^{-1}$]) $\sim 1$ in PIE or log($T/[K]) \sim 6$ in CIE.

 If we assume the typical picture that all these low-velocity warm absorbers (00-S1--S3, 04-S1--S5), are in PIE, as shown in Figure \ref{fig:colden} bottom 3 panels, they together span a wide range of gas outflow velocities, between 400 km s$^{-1}$ inflows to 400 km s$^{-1}$ outflows, and ionization parameters, between log($\xi$/[erg cm $s^{-1}$]) = $-$1-- 3.5. These all also exhibit similar column densities between log($N_H$/[cm$^{-2}$]) $\sim$ 20.5--21.5 and low line widths, $\sigma$ $\lesssim$ 150 km s$^{-1}$. We find the features are best represented by 5 physically distinct components in the 2004 dataset (04-S1, 04-S2, 04-S3, 04-S4, \& 04-S5), and by 3 in the 2000 dataset (00-S2 \& 00-S3), which is likely due to the difference in the depth of the data. 

\begin{figure}
    \label{itv}
    \centering
	\includegraphics[width=245pt]{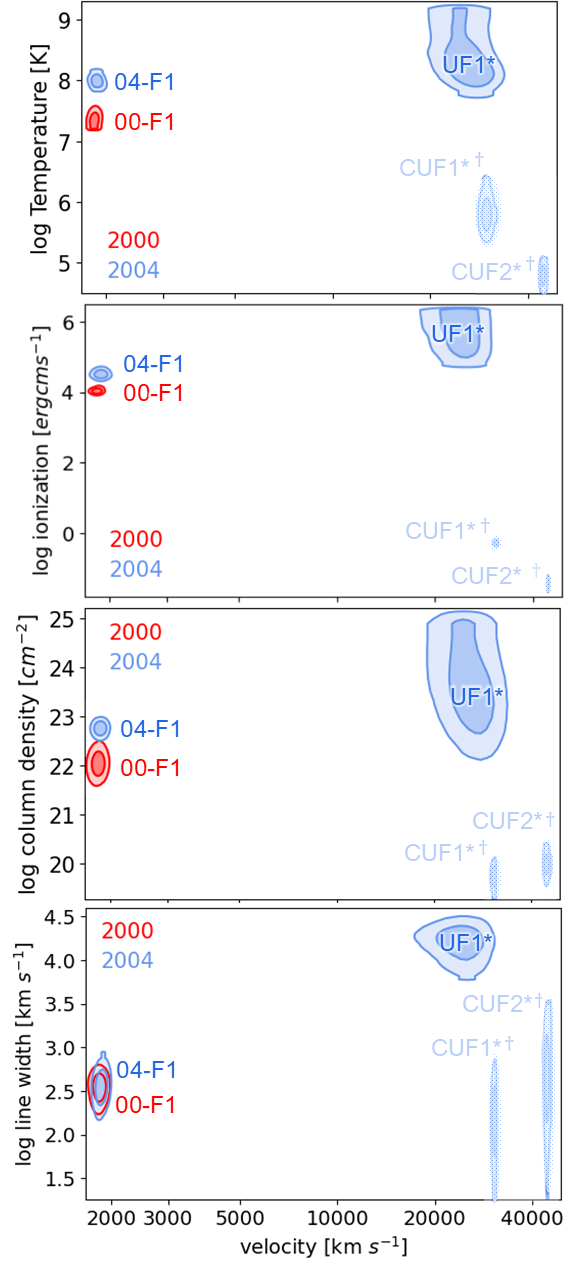}
    \caption{
    2D posterior probabilities of fast \& ultra-fast outflow (UFO) parameters with velocity on the x-axis. Temperature (top) for collisionally ionized (CIE) winds, ionization parameter for photoionized (PIE) winds, hydrogen column density, and line width (bottom) are on the y-axes. The 2000 results are red, and 2004 in light blue. Contours indicate 1 \& 2$\sigma$ errors. New hot \& likely unphysical cool UFOs are seen in the deeper 2004 data. The change of 00-F1/04-F1 may be an artifact of the time-averaged analysis.
    \\ \footnotesize{\emph{* ambiguous whether the CUF1-2 \& UF1 are CIE or PIE.}
    \\ \emph{$\dagger$ CUF1-2 are limited by systematic uncertainty in atomic data}.}
}
    \label{fig:posfast}
\end{figure}

\subsubsection{Fast Outflow (00-F1 \& 04-F1)} \label{res:fast}

We detect a moderately fast (v $\sim 1900$ km s$^{-1}$) and hot (log($T$/[K]) $\sim$ 8) ionized outflow (00-F1 \& 04-F1) best represented in CIE, with an increased $\Delta$DIC $\sim$ 20 (2004), compared to a similar absorber in PIE, near log($\xi$/[erg cm s$^{-1}$]) $\sim 4.5$, with further explanation for the preference and possible systematic uncertainties in \ref{fcie}. The line widths between epochs are consistent, while the column density and temperature  or ionization parameter appear to have increased (see Figure \ref{fig:posfast}). 
Simultaneously, the lower temperature CIE model is not favored in the 2000 data compared to a model in PIE with ionization log($\xi$/[erg cm s$^{-1}$]) $\sim 4.1$. The wind may have intrinsically changed in column density and temperature or ionization, but these measured differences are more likely attributed to the time-averaged nature of our approach and the lower signal-to-noise in the 2000 data, discussed further in Section \ref{fcie}. We also discuss potential physical interpretations for both PIE and CIE scenarios in Section \ref{fcie}.

We note that although the velocity shift of this absorber
is within $\sim$400 km s$^{-1}$ of our local frame of rest (the redshift of MCG-6 is equivalent to 2300 km s$^{-1}$), it is inconsistent at
$>$3$\sigma$. It is also unlikely to be a local absorber due to its high column density \citep{2005ApJ...631..733Y}, which is 2 orders of magnitude more than is usually reported
for the hot halo of the Milky Way, and its variability, as seen in XRISM \citep{2026ApJ..1003..103W}.

\subsubsection{Ultra-Fast Highly Ionized Outflow (UF1)}\label{res:UFO}

In the 2004 data, we detect a relativistic, hot absorber, `UF1', outflowing at $\sim8\%$ the speed of light, which was not previously detected in HETG grating data. We do not detect a similar absorber in the shorter 2000 data observation. 
For the detection in the 2004 data, both collisional and photoionized equilibrium gas models can fit the data equally well; see Section \ref{few} for further details.

Assuming it is a gas in CIE, this component has an outflow velocity of v $\sim$ 27,000 km s$^{-1}$, a high temperature of log($T$/[K]) $\sim$ 8, a large column density log($N_H$/[cm$^{-2}$]) $\sim$ 23.3, near the Compton-thick regime, and a very broad line width $\sigma$ $\sim$ 10,000 km s$^{-1}$. Alternatively, assuming it is a photoionized gas, it has a slightly lower outflow velocity of v $\sim$ 25,000 km s$^{-1}$, a high ionization parameter of log($\xi$/[erg cm s$^{-1}$]) $\sim$ 5.7, a higher column density of log($N_H$/[cm$^{-2}$]) $\sim$ 24, with a similarly broad line width $\sigma$ $\sim$ 10,000 km s$^{-1}$. See exact values in Table \ref{best_par}.
For discussion and limitations of this absorber detection and upper limits on the 2000 data counterpart, see Section \ref{diss}.

\begin{table} 
	\centering
	\caption{Best fit values for cold gas/dust in MCG-6-30-15 using \texttt{TBvarabs} \citep{2000ApJ...542..914W}, assuming metallic iron and neutral oxygen gas and dust, with 68\% credible intervals.}
	\label{tab:dust}
	\begin{tabular}{lccr} 
		\hline
         \hline
		\textbf{Model} \& Parameter & Unit & 2000 & 2004\\
        \hline
        \hline
         \bf{Iron} & & & \\
        \hline
        velocity $v$ & km s$^{-1}$ & 0$_{-100}^{+100}$ & -30$_{-60}^{+90}$  \\
        column density N$_{Fe}$ & $10^{17}$ cm$^{-2}$ & 3.0$_{-0.4}^{+0.2}$ & 2.4$_{-0.2}^{+0.3}$  \\
    	\hline
         \bf{Oxygen} (Option 1) & & & \\
        velocity $v$ & km s$^{-1}$ & \multicolumn{2}{c}{tied to Fe}  \\
        column density N$_{O}$ & $10^{17}$ cm$^{-2}$ & $<$ 1.0 & 0.62$_{-0.16}^{+0.06}$  \\
        N$_{O}$/N$_{Fe}$ Ratio & -- & $<$ 0.34 & 0.24$_{-0.05}^{+0.06}$\\
        \bf{Oxygen} (Option 2) & & & \\
        velocity $v$ & km s$^{-1}$ & -180$_{-300}^{+500}$ & 30$_{-160}^{+70}$  \\
        column density N$_{O}$ & $10^{17}$ cm$^{-2}$ & $<$ 1.4 & 1.3$_{-1.2}^{+0.6}$  \\
        N$_{O}$/N$_{Fe}$ Ratio & -- & $<$ 0.38 & 0.41$_{-0.2}^{+0.1}$\\
        \hline
	\end{tabular}
\end{table}

\begin{table} 
	\centering
	\caption{Local neutral \citep[\texttt{TBabs},][]{2000ApJ...542..914W} and Collisionally Ionized Hot Milky Way Halo absorption with 68\% credible intervals.}
	\begin{tabular}{lccr} 
		\hline
         \hline
		\textbf{Model} \& Parameter & Unit & 2000 & 2004 \\
        \hline
        \hline
         \multicolumn{2}{l}{\bf{Neutral Milky Way Absorption}}\\
        \hline
        Velocity $v$ & km s$^{-1}$ & 120$_{-75}^{+35}$ & 60$_{-50}^{+30}$  \\
        column density $N_H$ & $10^{20}$cm$^{-2}$ &3$_{-1}^{+2}$ &4.4$_{-1.2}^{+0.7}$\\
                \hline
         \multicolumn{2}{l}{\bf{Hot Milky Way Halo}} \\
		\hline
        Velocity $v$ & km s$^{-1}$ &-50$_{-100}^{+100}$ & 25$_{-60}^{+30}$ \\
        log Temperature $T$& K & 6.2$_{-0.1}^{+0.1}$ & 6.0$_{-0.1}^{+0.1}$  \\
        log Column Density $N_H$ & cm$^{-2}$ &20.0$_{-0.3}^{+0.3}$ &19.6$_{-0.1}^{+0.1}$\\
		  Line Width $\sigma$ & km s$^{-1}$ &$<$ 200 & 120$_{-50}^{+60}$  \\
        \hline
	\end{tabular}
    \label{tab:MW}
\end{table}

\begin{table*} 
	\centering
	\caption{Detected moderate inflows (negative) \& outflows (positive) of MCG-6-30-15 moving at $<$1000 km s$^{-1}$ listed at the top, fast outflows moving $>$1000 km s$^{-1}$ listed in the middle, and likely unphysical cool relativistic absorbers listed at the bottom. The columns list the preferred ionization process: photoionization equilibrium models (PIE), or collisionally ionized equilibrium models (CIE), and best-fit values with 68\% credible intervals.}
	\label{tab:best}

\begin{tabular}{lcccccccr}
        \hline
        \hline
        \multicolumn{9}{c}{\textbf{Slow Absorbers $<$1000 km s$^{-1}$}} \\
        \hline
        \hline
        Name & Pref. & Significance & velocity & temperature & ionization & column density & line width & Note \\
        & Process & $\Delta DIC$ & $v$ [km s$^{-1}$] & log$T$/[K] & log$\xi$/[erg cm s$^{-1}$] & log$N_H$/[cm$^{-2}$] & $\sigma$ [km s$^{-1}$] & \\
        \hline
        \multicolumn{9}{c}{2000 Dataset}\\
        \hline
        00-S1  & \textbf{PIE} & \textbf{623} &  250$_{-10}^{+20}$ &  -- &  3.0$_{-0.02}^{+0.07}$ & 21.54$_{-0.05}^{+0.05}$ &  100$_{-10}^{+20}$ & \hphantom{{\footnotesize by 1 feature}}\\
          \\
         00-S2 & \textbf{PIE} & \textbf{110} & 20$_{-30}^{+30}$ &  -- & 1.83$_{-0.04}^{+0.1}$& 21.10$_{-0.02}^{+0.1}$ &  $<$60\\
         &\textbf{CIE} & \textbf{120} & 30$_{-30}^{+30}$ &  6.07$_{-0.02}^{+0.01}$ & -- & 21.16$_{-0.03}^{+0.07}$ &  $<$50\
         \\
         \\
         00-S3 & \textbf{PIE} & \textbf{45} & -140$_{-90}^{+30}$ &  -- & -0.2$_{-0.2}^{+0.2}$ & 20.8$_{-0.2}^{+0.1}$ &  $<$500\\
        \hline
        \multicolumn{9}{c}{2004 Dataset} \\
        \hline
04-S1 & \textbf{PIE} & \textbf{1922} & 220$_{-10}^{+20}$ & -- & 2.94$_{-0.07}^{+0.05}$ & 21.3$_{-0.2}^{+0.1}$ & 120$_{-20}^{+10}$ &  \\
\\
04-S2 & \textbf{PIE} & \textbf{487} & -20$_{-10}^{+20}$ & -- & 2.01$_{-0.07}^{+0.04}$ & 21.19$_{-0.05}^{+0.05}$ & $<$ 40 &  \\
& \textbf{CIE} & \textbf{490} & 10$_{-20}^{+10}$ & 6.06$_{-0.01}^{+0.01}$ & -- & 21.28$_{-0.07}^{+0.02}$ &  50$_{-10}^{+10}$\\
\\
04-S3 & \textbf{PIE} & \textbf{113} & -350$_{-40}^{+40}$ & -- & -0.21$_{-0.08}^{+0.06}$ & 20.85$_{-0.09}^{+0.05}$ & $<$ 30 &  \\
\\
04-S4 & \textbf{PIE} & \textbf{11} & 280$_{-30}^{+30}$ & -- & 3.4$_{-0.1}^{+0.1}$ & 21.1$_{-0.2}^{+0.1}$ & 90$_{-10}^{+50}$ & \\
\\
04-S5 & \textbf{PIE} & \textbf{30} & -60$_{-60}^{+20}$ & -- & 1.1$_{-0.2}^{+0.2}$ & 20.6$_{-0.2}^{+0.1}$ & 80$_{-20}^{+30}$ &  \\
& \textbf{CIE} & \textbf{28} & -70$_{-20}^{+70}$ & 5.8$_{-0.1}^{+0.2}$ & -- & 20.6$_{-0.1}^{+0.2}$ & 90$_{-30}^{+20}$ &  
\\
        \\
        \hline
        \hline
        \multicolumn{9}{c}{\textbf{Fast Absorbers $>$1000 km s$^{-1}$}} \\
        \hline
        \hline
        Name & Pref. & Significance & velocity & temperature & ionization & column density & line width & Note \\
        & Process & $\Delta DIC$ & $v$ [km s$^{-1}$] & log$T$/[K] & log$\xi$/[erg cm s$^{-1}$] & log$N_H$/[c$m^{-2}$] & $\sigma$ [km s$^{-1}$] & \\
        \hline
        \multicolumn{9}{c}{2000 Dataset}\\
        \hline
        00-F1          & \textbf{PIE}  & \textbf{213} & 1840$_{-30}^{+40}$ &  -- & 4.1$_{-0.2}^{+0.2}$ & 22.0$_{-0.1}^{+0.1}$ &  390$_{-50}^{+100}$\\
        & \textbf{CIE}  & \textbf{184} & 1850$_{-70}^{+70}$ &  7.4$_{-0.07}^{+0.2}$ & -- & 21.1$_{-0.1}^{+0.05}$ &  360$_{-70}^{+70}$\\
        \hline
        \multicolumn{9}{c}{2004 Dataset} \\
        \hline
04-F1 & PIE & 243 & 1880$_{-50}^{+50}$ & -- & 4.5$_{-0.1}^{+0.1}$ & 22.4$_{-0.1}^{+0.1}$ & 550$_{-90}^{+80}$ &  \\
& \textbf{CIE} & \textbf{263} & 1890$_{-20}^{+60}$ & 7.97$_{-0.05}^{+0.05}$ & -- & 22.79$_{-0.09}^{+0.06}$ & 360$_{-90}^{+80}$ &  \\
\\
UF1 & \textbf{PIE} & \textbf{16} & 25000$_{-2000}^{+1000}$ & -- & 5.8$_{-0.6}^{+0.3}$ & 24.0$_{-0.6}^{+0.1}$ & 16000$_{-4000}^{+4000}$ &  \\
 & \textbf{CIE} & \textbf{25} & 27000$_{-3000}^{+1000}$ & 8.1$_{-0.1}^{+0.3}$ & -- & 23.2$_{-0.1}^{+0.5}$ & 11000$_{-1000}^{+5000}$ &  \\
        \end{tabular}
\begin{tabular}{lcccccccr}
        \hline
        \hline
        \multicolumn{9}{c}{\textbf{Components Limited by Systematic Uncertainty in Atomic Data}} \\
        \hline
        \hline
        Name & Pref. & Significance & velocity & temperature & ionization & column density & line width & Note \\
        & Process & $\Delta DIC$ & $v$ [km s$^{-1}$] & log$T$/[K] & log$\xi$/[erg cm s$^{-1}$] & log$N_H$/[c$m^{-2}$] & $\sigma$ [km s$^{-1}$] & \\
        \hline
        \multicolumn{9}{c}{2004 Dataset}\\ 
        \hline
CUF1 & \textbf{PIE} & \textbf{19} & 32100$_{-100}^{+100}$ & -- & -0.2$_{-0.2}^{+0.1}$ & 19.8$_{-0.2}^{+0.1}$ & 700$_{-100}^{+200}$ &  {\footnotesize *Constrained} \\
  & \textbf{CIE} & \textbf{15} & 29800$_{-100}^{+100}$ & 5.8$_{-0.1}^{+0.2}$ & -- & 19.6$_{-0.2}^{+0.2}$ & 600$_{-200}^{+200}$ &  {\footnotesize by 3 features}\\
\\
CUF2 & \textbf{PIE} & \textbf{36} & 44600$_{-100}^{+100}$ & -- & -1.4$_{-0.1}^{+0.1}$ & 20.0$_{-0.1}^{+0.1}$ & 600$_{-300}^{+500}$ &   {\footnotesize *Constrained}\\
 & \textbf{CIE} & \textbf{45} & 48000$_{-100}^{+100}$ & 4.8$_{-0.1}^{+0.1}$ & -- & 20.0$_{-0.1}^{+0.1}$ & 400$_{-200}^{+300}$ &  {\footnotesize by 1 feature}\\
\\
        \hline
\end{tabular}
    \label{best_par}
\end{table*}

\begin{figure*}
    \centering
	\includegraphics[width=\textwidth]{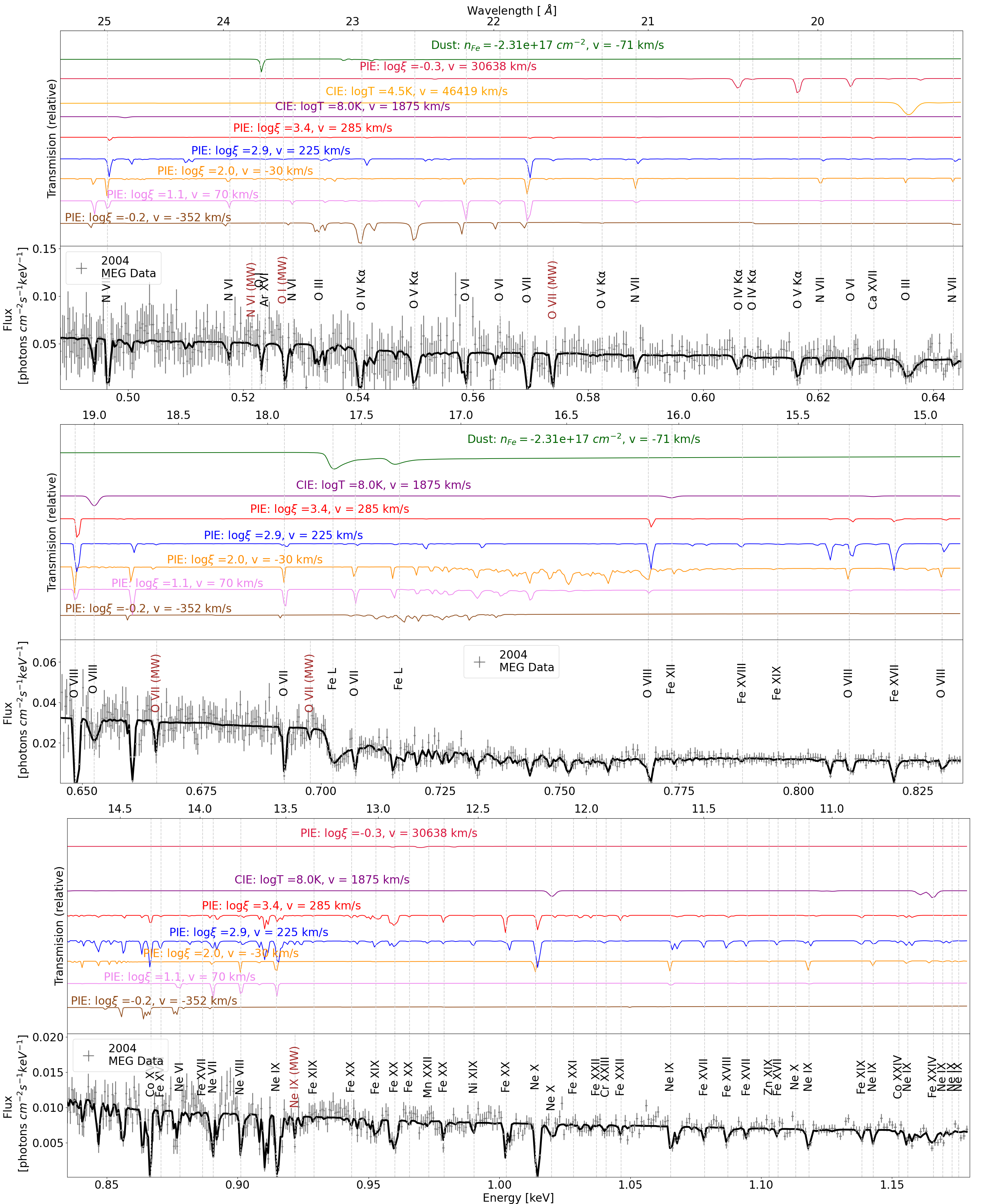}
    \caption{Lower panels: Multidimensional best fit model (solid black line) with 2004 unfolded MEG data (gray points), and ions contributing the most absorption at a given energy. Upper panels: The transmission for 10 absorbers intrinsic to MCG-6-30-15 (various colors). Features from neutral Galactic and hot Milky Way Halo absorption are labeled `MW'. Spectra were combined and optimally binned for plotting purposes.}
    \label{fig:lowspec2004}
\end{figure*}
\begin{figure*}
    \ContinuedFloat
	\includegraphics[width=\textwidth]{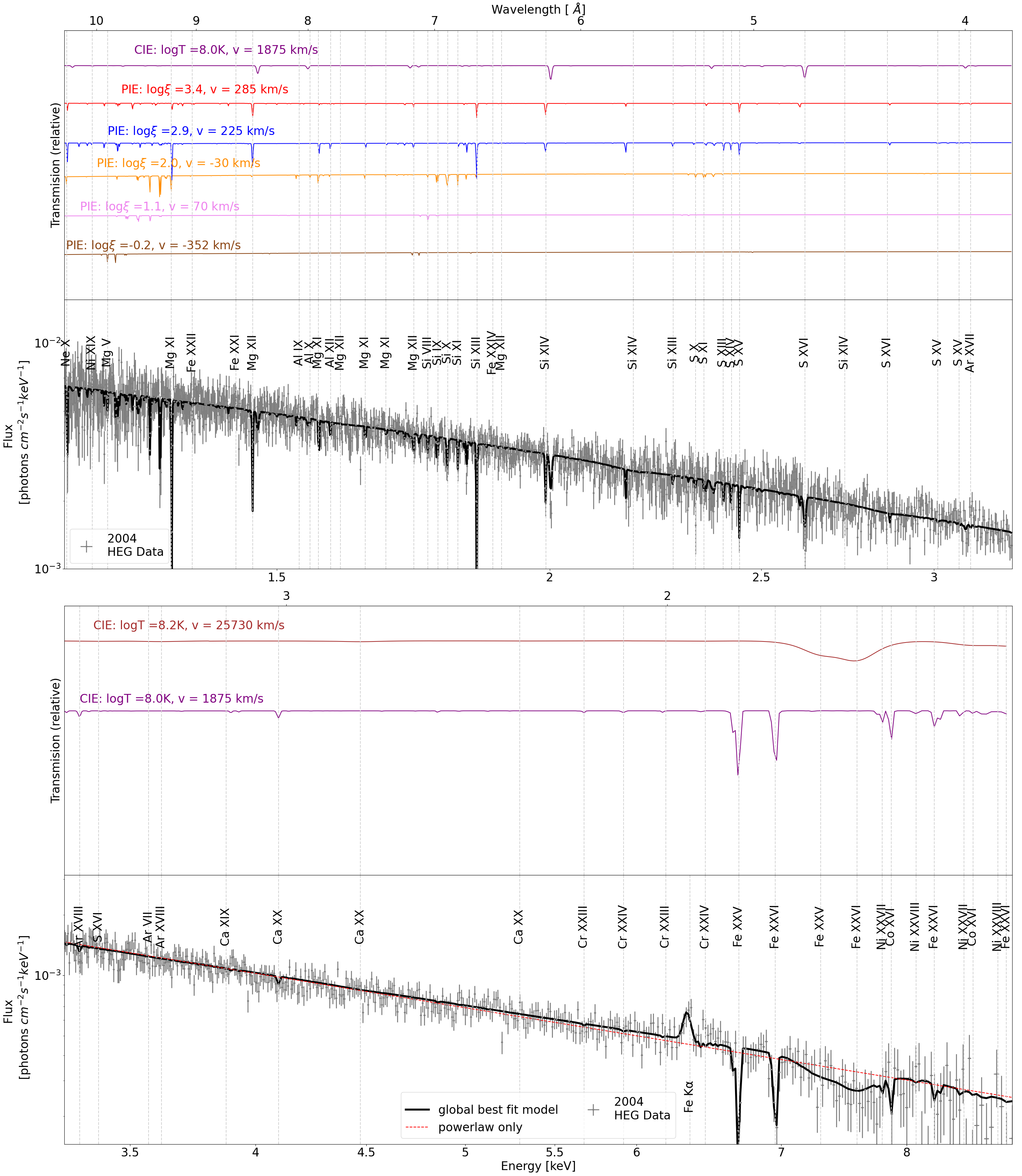}
    \caption{
    (Continued) Lower panels: Multidimensional best fit model (black line) with unfolded 2004 HEG data keV (gray points) in log-log space. The last lower panel includes the powerlaw component (dashed red line) to highlight the broad hot UFO component $\sim$7--8 keV, which spans $\sim$ 30 data points dipping below the main continuum.}
    \label{fig:highspec2004}
\end{figure*}

\begin{figure*} 
	\includegraphics[width=\textwidth]{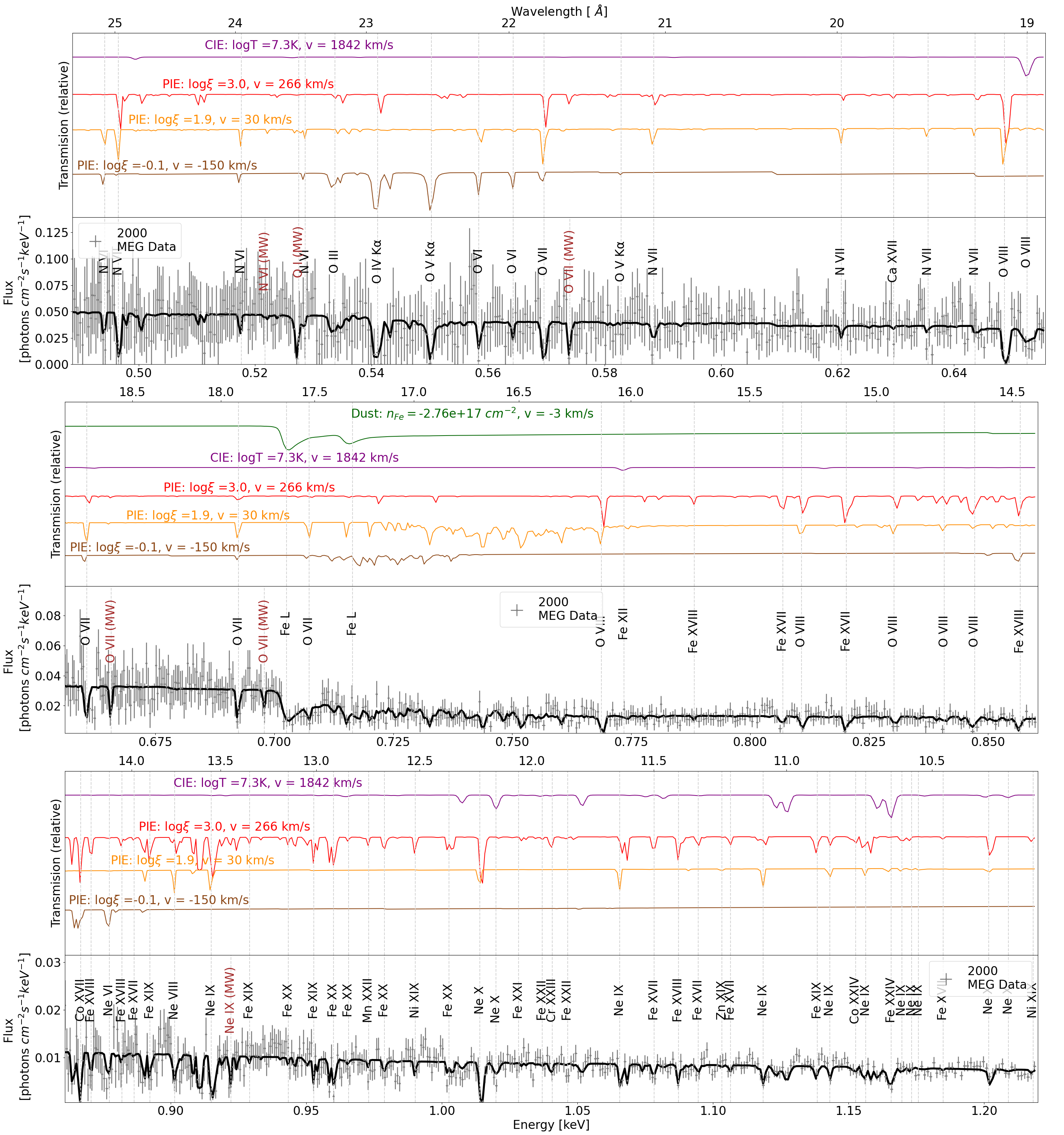}
    \caption{Lower panels: Multidimensional best fit model of 2000 (solid black line) MEG data (gray points), and ions contributing the most absorption at a given energy. Upper panels: The transmission for 5 absorbers intrinsic to MCG-6-30-15 (various colors). Features from neutral Galactic and hot Milky Way Halo absorption are labeled `MW'. Spectra were combined and optimally binned for plotting purposes.}
    \label{fig:lowspec2000}
\end{figure*}

\begin{figure*} 
    \ContinuedFloat
	\includegraphics[width=\textwidth]{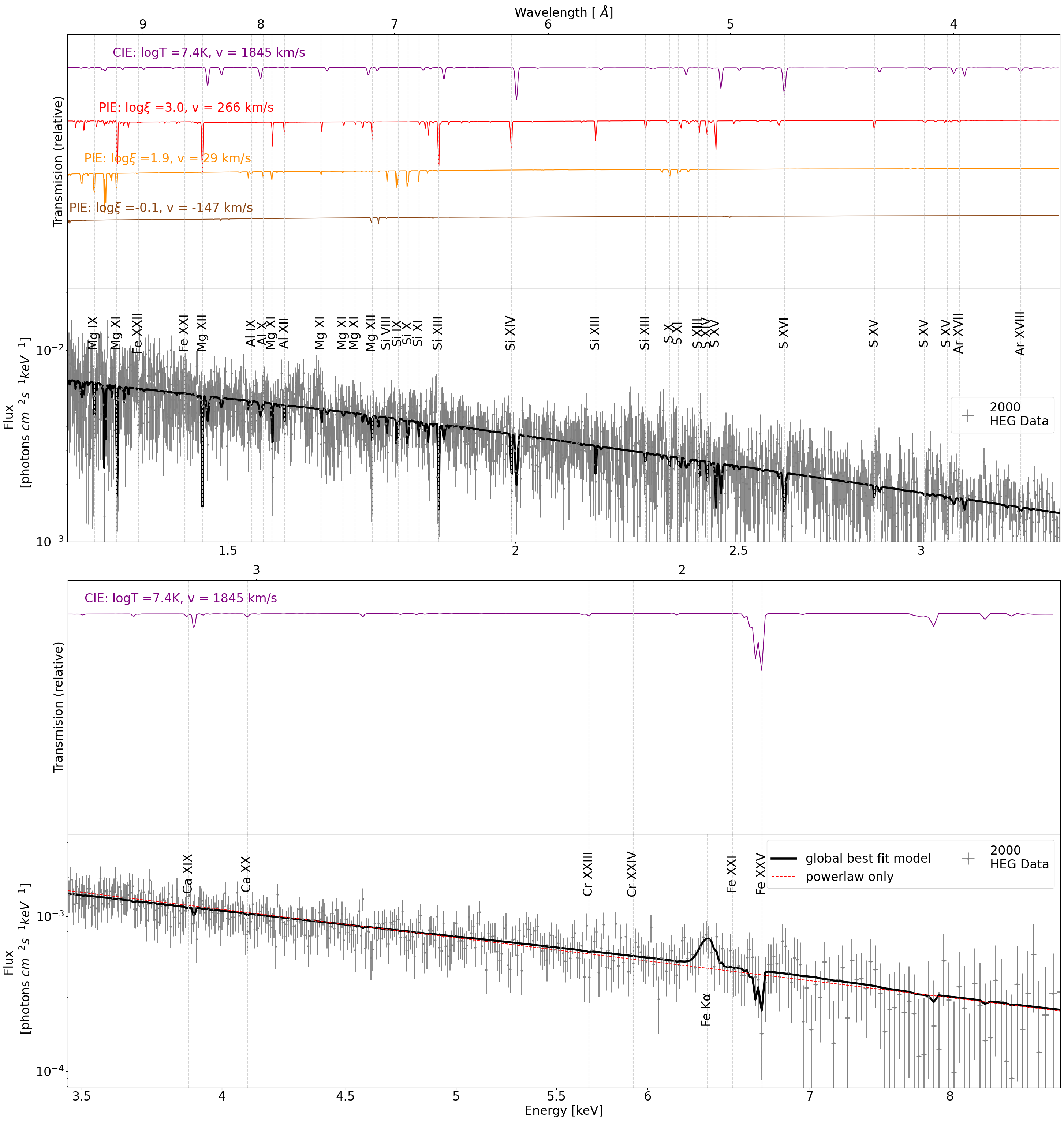}
    \caption{(Continued) Lower panels: Multidimensional best fit model (black line) with 2000 HEG data keV (gray points) in log-log space. The last lower panel includes the powerlaw component (dashed red line) to highlight the broad hot UFO component was undetected in the 2000 data.}
    \label{fig:specspec2000}
\end{figure*}

\subsubsection{Cold Gas and Dust in MCG-6} \label{res:dust}
We find that among the top 4 most statistically significant absorption components in both observations is a model for Fe L absorption edges. Our model assumes metallic iron dust given the high dust extinction in this source \citep{1997MNRAS.291..403R}. We note, however, that we do not test models for Fe I, Fe II, Fe III gas or iron dust in other molecules, which may equally well represent the same edge features, but be indistinguishable with this data set \citep{2021ApJ...908...52S}. 

Assuming metallic iron dust using \emph{TBvarabs} model with free iron and/or oxygen column densities, and therefore free relative abundances of  O:Fe, is strongly preferred (difference in $\Delta$DIC $>$ 150) over a \emph{TBabs} model in MCG-6, which uses fixed abundance ratios measured for our Galaxy. This means that the O:Fe ratio we observe in MCG-6 is strongly disfavored to resemble the O:Fe in our Galactic ISM.

Assuming all the iron is in metallic iron dust, we measure the iron column density to be $N_{Fe} \sim$ 2.4--3 $\times 10^{17}$ cm$^{-2}$, consistent within 3$\sigma$ between 2000 and 2004.
We find the iron dust has a line of sight velocity consistent within 1$\sigma$ of the rest frame of MCG-6. We note, however, there are large systematic uncertainties on the velocity (up to $\sim$100s km s$^{-1}$), depending on the model used \citep{2024ApJ...965..172C}, the neutral gas, the low-ionization gas, or dust molecule assumed \citep{2021ApJ...908...52S}, and taking into account the absolute energy resolution of HETG is near $\sim$100 km s$^{-1}$, \citep{2005PASP..117.1144C}. Including these effects, if the dust is embedded within a wind, it could have a velocity consistent with any of the warm absorbing systems found within this object.

Assuming there is also neutral oxygen gas and/or dust in MCG-6, using the \emph{tbvarabs} model, the oxygen has a column density of N$_O$ $\sim$ 0.6-1.3 $\times 10^{17}$ in 2004 and $<$ 1.0--1.4 $\times 10^{17}$ in 2000, depending on whether we assume the Fe originates from matter with the same velocity, and either tie the velocity of O to Fe or not. This implies O:Fe ratios $\sim$0.25-0.4 or less, which is inconsistent with solar abundance ratios by 2 orders of magnitude \citep[O:Fe $\sim$20,][]{1998SSRv...85..161G}. See Table \ref{tab:dust} for exact values for both options.

\subsubsection{Neutral \& Hot Milky Way Absorption} \label{res:NMW}

To utilize the full spectral resolution of the data and avoid biasing our measurement, we keep both the hydrogen column density (via \emph{Tbabs}) and velocity shift (via \emph{zmshift}) of neutral galactic absorption free throughout our modeling process. We find the features are consistent with our local frame of rest within 2$\sigma$ statistical uncertainty, and within the absolute energy scale calibration for HETG \citep{2005PASP..117.1144C}. 

We find a hydrogen column density of $N_H \sim$4 $\times10^{20}$ cm$^{-2}$ is consistent within 1$\sigma$ for both datasets and the reported values of Galactic absorption in the direction of our target galaxy (3.6 $\times10^{20}$ cm$^{-2}$, \citealt{2016A&A...594A.116H}). 

In both the 2000 and 2004 datasets, we find another local hot, collisionally ionized absorber log($T$/[K]) $\sim$ 6 representing X-ray absorption in the warm-hot halo of our Galaxy (\citealt{2012ApJ...756L...8G}, \citealt{2021ApJ...918...83D, 2024A&A...681A..78L}, and references therein) for both observations. See Figure \ref{fig:MWhot} for 2D posterior probabilities and Table \ref{tab:MW} for exact values. 

\begin{figure}
    \subfloat[\centering ]{{\includegraphics[width=\linewidth]{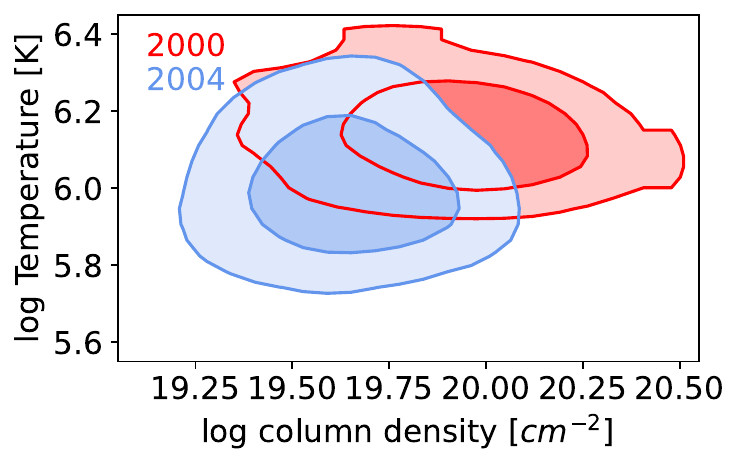} }}%
    \qquad
    \subfloat[\centering]{{\includegraphics[width=.95\linewidth]{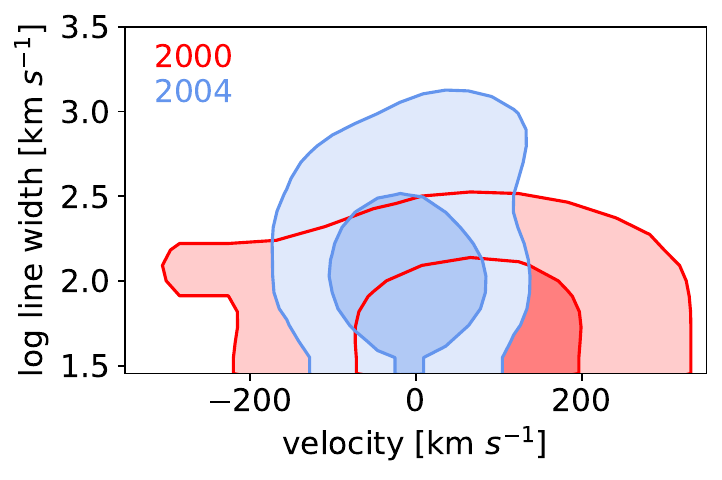} }}%
    \caption{2D posterior probabilities of collisionally ionized Milky Way Hot CGM/Halo absorber parameters temperature vs outflow velocity (a) and line width vs hydrogen column density (b). The 2000 results are dark red, and the 2004 results are light blue, with contours indicating 1 \& 2$\sigma$ errors, which are consistent between both datasets.}
    \label{fig:MWhot}
\end{figure}

\vspace{5mm}
\subsubsection{Components Limited by Systematic Uncertainty in Atomic Data (CUF1 \& CUF2)}\label{sec:pot}
There are $\sim$4 significant features unaccounted for below 1 keV in the 2004 dataset by our other models. 
Our framework finds that these features are two cool or low-ionization parameter (low-IP) components, both ambiguously either in CIE or PIE, traveling at relativistic speeds of $0.1$--$0.15c$; however, a simpler explanation is that these lines are higher transition O IV and OV lines, \citep[as ][suggests]{2010ApJ...708..981H} an interpretation supported by theoretical calculations and some experimental measurements, and which may be missing from our warm absorber models. We discuss the limitations of interpreting results such as these due to systematic uncertainties in atomic databases in Section \ref{few}.

\section{Discussion}\label{diss}
Our work finds good agreement with previous studies, independently supporting all previously detected wind component velocities, while increasing precision in measured gas parameters, expanding upon their conclusions, and making new detections. We see the WAs in new, unprecedented detail, uncovering a previously unseen potential velocity-ionization trend. We detect multiple winds as equally well or better represented in CIE for the first time, two stable and one potentially varying. We find a hot UFO not previously detected in HETG. Our conclusions on Fe L-shell absorption, which for the first time include the deeper 2004 HETG data, also support previous work that solid Fe is likely embedded within a wind in MCG-6, as opposed to the host galaxy ISM.

\subsection{Comparison of slow absorbers to previous works}

Most previous works find 1--2 intrinsic absorbers with ionization around log($\xi$/[erg cm s$^{-1}$]) $\sim$ 0--2, moving at velocities between 100--200 km s$^{-1}$, \citep[e.g.][]{2010ApJ...708..981H, 2001ApJ...554L..13L}.
Our work finds 3 distinct low-ionization, low-velocity absorbers in the 2000 data, and 5 in the deeper 2004 data. Our new detections are likely due to our framework, which uses an agnostic, standardized statistical stopping point for both components in CIE and PIE, which is especially advantageous when components like these overlap for many absorption features. We are also able to measure each of these components' properties with higher precision and accuracy using our finely sampled gas parameters and novelly including line width as a free parameter.

Among these slow absorbers, we find a wider range of velocities and ionization parameters, and we are the first to report potential inflows.
Our analysis finds lower column densities for each of the slow components compared to previous works. This is likely a consequence of these gases absorbing similar ions, splitting the column densities between a greater number of components (see Figure \ref{fig:lowspec2004} and \ref{fig:lowspec2000}, for example, ions O VII, O VIII, Ne IX, Ne X). 

\begin{figure*}%
    \centering
    \subfloat[\centering slow absorbers]{{\includegraphics[width=.46\linewidth]{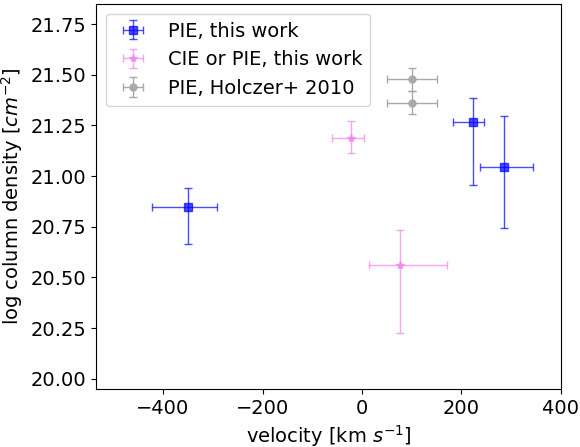} }}%
    \qquad
    \subfloat[\centering fast absorbers]{{\includegraphics[width=.46\linewidth]{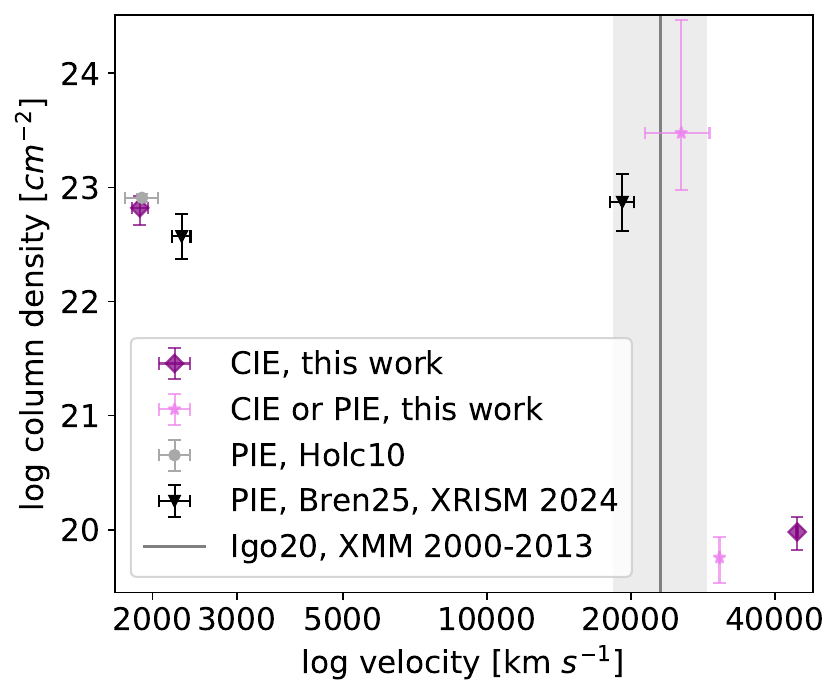} }}%
    \caption{Column density vs outflow velocity for slow ($<$1000 km s$^{-1}$, left) and fast absorbers ($>$1000 km s$^{-1}$, right) detected in the 2004 data by our work, with blue squares for photoionized equilibrium (PIE), purple diamonds for collisionally ionized equilibrium (CIE), and pink stars for ambiguous PIE or CIE preference. We compare to \citet[][: Holc10]{2010ApJ...708..981H} results (gray circle) of the same dataset, and from different datasets (\citealt[][XRISM, black triangle: Bren25]{2025ApJ...995..200B}; \citealt[][XMM, gray line: Igo20]{2020MNRAS.493.1088I}). Error bars represent 90\% confidence intervals. Our work finds more overall slow absorbers with low column densities, likely because they share absorption for some of the same ionic transitions. We detect many more relativistic components, varying from hot to cool.}%
    \label{fig:dissfast}%
\end{figure*}
\subsection{First CIE modeling of slow absorbers in MCG-6 with CGM-like properties} \label{sec:cvp}
Among the slow absorbers, we report the first potential collisionally ionized absorbers in MCG-6.
The statistical preference for a PIE or CIE can arise because each model produces different ion fractions. PIE models have smoother distributions across ionization parameter, while ions in CIE can exist in more narrow temperature windows, peaking in ion fractions more sharply, and for certain ions, peaking at much higher ion fractions \citep[e.g., OVII, Ne IX, Fe XXV]{1999ApJ...512..184N}. If both models can equally well model the data, that could indicate a regime in which these differences in ion fractions are negligible, and/or indistinguishable by the data.

We find two such cases among the slow warm-absorbing zones of MCG-6, in which 00-S2, 00-S2, and 04-S5 can be well represented by either a PIE or CIE model. Since these are among a range of absorbers preferred in PIE which are similar in properties, it would be plausible to assume these absorbers are in PIE as well, supporting a typical scenario of these winds.

Alternatively, these absorbers, especially 00-S2/00-S2, if assumed to be in CIE, have a temperature (log($T/[K]) \sim 6$), line width ($<$100 ), and outflow velocity ($\sim$0). These parameters can imply non-AGN origin of this component, such as the galactic and cricum-galactic (CGM) absorption in MCG-6. The column density 00-S2/00-S2, however, is notably more than a magnitude higher than typically measured for our Galaxy, which is comparable in mass to MCG-6. This could potentially be explained by the difference in our line of sight through the target galaxy, which may include more of the MCG-6 galactic disk and ISM, and in our own Galaxy, it is known to vary by sightline \citep{2018MNRAS.474..696G}. Or the similarities to the galactic and ISM absorption could be coincidental, and these absorbers could be collisional due to shock heating by other winds interacting with the ISM.
Since absorbers in CIE have not been considered in most sources with winds, this could be an early indication of a new, common subtype of X-ray absorbers. 
More detections to constrain their overall occurrence and range in measured parameters could help determine their nature.

\begin{figure}%
    \includegraphics[width=\columnwidth]{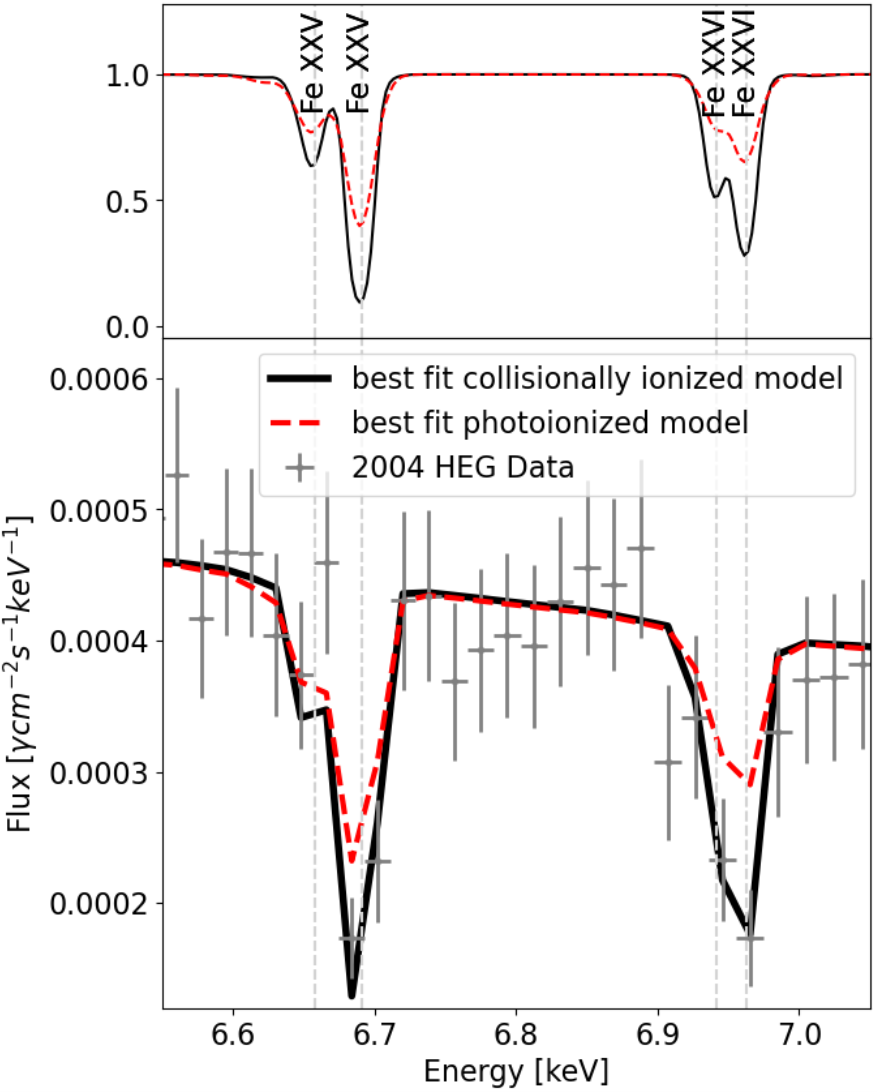}
    \caption{(bottom) A portion of the 2004 HEG spectrum overplotted with two potential models, with flux on the y-axis and energy in keV on the x-axis.  (top) The  transmission spectrum of the two absorption models with  their relevant ions. The $\sim$1900 km $s^{-1}$ collisionally ionized absorber (solid black line, component 04-F1) produces a greater relative depth in Fe lines compared to lines from other ions, which yields an overall greater change in $\Delta$DIC, compared to a similar photoionized model (dashed red line). For more details, see Section \ref{fcie}}%
    \label{fig:fe2425}%
\end{figure}

\subsection{Comparison of Fast CIE Absorber}\label{fcie}
Previous works on MCG-6 \citep[e.g.][]{2010ApJ...708..981H, 2006ESASP.604..475Y} find a narrow, fast $\sim$1900 km s$^{-1}$, high-ionization component (log($\xi$/[erg cm s$^{-1}$]) $\sim$ 4), with a high column density around log($N_H$/[cm$^{-2}$]) $\sim$ 22.7 (00-F1/04-F1). While previous detections assume it's in PIE, we find CIE is better able to represent the component  in the deeper 2004 dataset. The preference is largely driven by the increased relative  shape and column depth of various ions, especially FeXXV and FeXXVI (see in Figure \ref{fig:fe2425}) relative to ions at lower energies, \citep[][]{1999ApJ...512..184N}. We note that we assume solar abundances, and non-solar abundances may affect the fit, but such analysis is beyond the scope of this work.

This absorber appears to have increased in column density and temperature by an order of magnitude from 2000 to 2004, with a significance of $>$3$\sigma$. However, this may also be an artifact of the time-averaged analysis, or the assumption of ionization equilibrium, since this absorber likely varies on much shorter timescales \citep{2026ApJ..1003..103W}. 
Furthermore, the ionization is largely constrained by the presence of the FeXXVI line, which is especially weak in the 2000 data (See Panel 2 of Figure 8).

Since it is possible that the statistical preference for a model in CIE can be encompassed in systematic uncertainties, we consider both CIE and PIE possibilities. A component in PIE is the standard assumption, indicating an absorber ionized by the AGN's radiation, and potentially very close to the central engine due to its high ionization and velocity. 

An absorber near the AGN with a spectrum resembling that of collisional ionization is more complex to explain. Near the AGN, a CIE absorber is unlikely to be dominated by the physics of typical low-density collisional plasmas, and may have more similarities with a high-density non-relativistic plasma, with free-free heating dominating over Compton heating, resulting in altered ion fractions \citep{2008A&A...487..895R}. Alternatively, if farther from the AGN, a fast-moving CIE component could be explained by shocks interacting with the ISM outside the BLR.

\subsection{UFO Confirmed by \emph{Chandra's} HETG}\label{uf}
Independent of this work, analysis of MCG-6 data from XRISM also detects a broad 0.08c UFO \citep{2025ApJ...995..200B}, assumed to be in PIE. No previous works analyzing MCG-6 with \emph{Chandra} HETG have reported a hot UFO, while our methodology finds it midway through our analysis framework, as the 6th out of 11 statistically significant absorbing components in the 2004 data, and ambiguously either in PIE or CIE. The only previous (pre-XRISM) work to report a UFO between \emph{Chandra} or XMM-Newton is \citet{2020MNRAS.493.1088I}, which applied a different methodology to XMM EPIC-pn (CCD) data, searching specifically for variable H- and He- like ion lines. \citet{2020MNRAS.493.1088I} similarly find a relativistic outflow at $\sim$0.08c, constrained by an FeXXV line at 7.26 keV, but do not report other measurable properties to compare. Using a \texttt{CLOUDY} model, we constrain its ionization parameter or temperature, its near Compton-thick column density (see Figure \ref{fig:dissfast}), and broad line width. We identify the broadened features as a mixture of FeXXV and FeXXVI, shown in Figure \ref{fig:highspec2004}.

While XRISM has a higher sensitivity in the FeK band, this highlights that the deep Chandra HETG observations are also capable of detecting UFOs and can be used to establish a long-term variability baseline (Ogorzalek et al. in prep). 

We note, however, that broad absorption in this 7 keV region is highly sensitive to the spectral features of relativistically broadened reflection, which is not well constrained by HETG alone. The recent joint XRISM Resolve, XMM-Newton, and NuSTAR campaigns with better hard-X-ray coverage and time-resolved treatment of relativistic reflection \citep[][]{2025ApJ...995..200B, 2026ApJ..1003..103W} report different reflection parameters, such as disk inclination. We checked whether reflection, particularly disk inclination, which affects the blue wing of the relativistic line, affects the detection or measurements of the hot UFO. Using the \citet{2025ApJ...995..200B} best-fit value, we find that our reported outflow velocity may be systematically biased by $\sim$10\%, which is below 1$\sigma$. Importantly, the hot UFO is detected irrespective of the reflection parameters.

The hot UFO is not detected in the 2000 data, potentially due to the shorter observation time. Assuming the 2000 epoch contains an absorber exhibiting properties within 1$\sigma$ of those measured in 2004, we find an upper limit of $<$10$^{22.9}$ cm$^{-2}$ if in CIE and $<$10$^{23.5}$ cm$^{-2}$ if in PIE. 

Overall, its important to highlight that \emph{with sufficient observation time and signal-to-noise, HETG is capable of detecting high-ionization broad UFO absorbers.}

\subsection{Comparison of detected cold Fe to other works}
Our work supports the detection of Fe L shell edge at 0.7 keV, likely caused by neutral iron gas/dust in MCG-6, similar to the conclusions of \citealt{2001ApJ...554L..13L}. We, in particular, infer an O-poor solid Fe compound, like \citealt{2003ApJ...596..114S}. The presence of O I gas is also evident from the 23.7 \r{A} (0.523 keV) line feature. We apply models assuming metallic Fe dust and O I gas, and the total abundance of O (gas plus dust) is constrained by the continuum absorption of the \emph{tbvarabs} model. Therefore, we take our measured O column as a measure of the total cold-phase O in our sight line. If we assume that the Fe L and the O K absorption arises from the same material, this material would have an O:Fe ratio $\leq$ 0.25 which is 
incompatible with existing models of ISM dust containing Fe \citep[e.g. O:Fe $\sim$4:1,][]{1996NASCP3343...87S}, or the O:Fe ratio 2:1 modeled in MCG-6 by \citet{2001ApJ...554L..13L}. We conclude that a large fraction of the Fe column is from an O-poor solid compound, like metallic Fe, surviving in a warm ionized AGN wind where O atoms cannot remain neutral (as in the diffuse ISM).

The velocity of the feature is uncertain due to a variety of systematic effects of the Fe dust model; however, since we measure it in the rest frame of MCG-6, it is likely still within $\sim$100s km s$^{-2}$ of that value. Therefore, our analysis finds many potential slow-moving warm absorbers within which the dust could be embedded. This conclusion would support the mechanism of radiation pressure to drive WAs, since dust can enhance the effectiveness of thrust on the wind \citep{2008MNRAS.385L..43F, 2018MNRAS.476..512I}. 

\begin{figure}
    \subfloat[\centering]{{\includegraphics[width=\columnwidth]{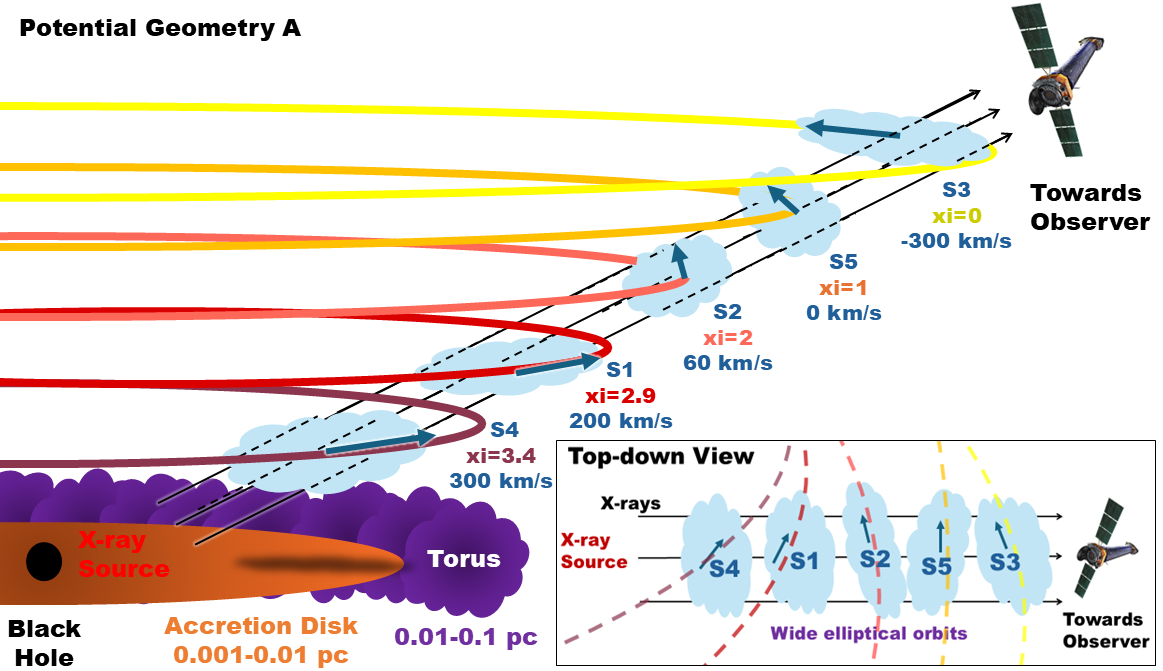} }}%
    \qquad
    \subfloat[\centering]{{\includegraphics[width=\columnwidth]{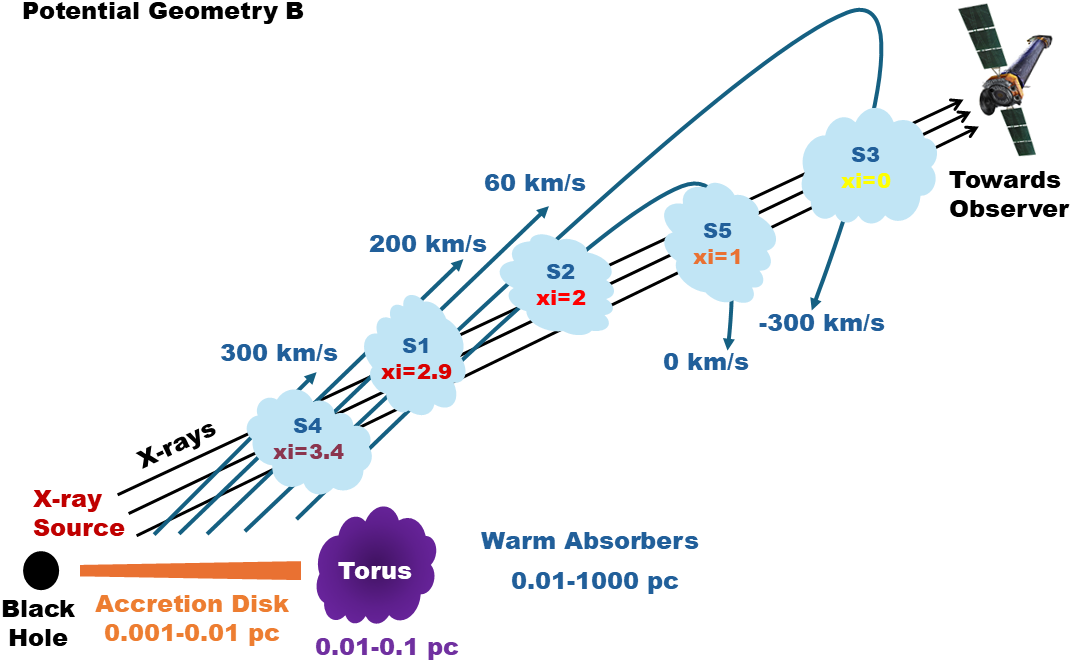} }}%
    \caption{Potential geometric orientations for slow winds, assuming similar densities: (a) 3D and top-down view of gases in a wide out-spiraling elliptical orbits (b) edge-on view of gases at different distances from the accretion disk (scaling specific to MCG-6, a 10$^6 M_\odot$ SMBH), with inflows falling back towards the black hole, potentially originating from the same outflowing gas.}
    \label{fig:geos}
\end{figure}

\subsection{Potential Origin and Geometry of absorbers}
Among the slow absorbers, there is a  $\sim$3$\sigma$ linear correlation between velocity and ionization. Components with a potential inflow ($\sim$400 km s$^{-1}$) have lower ionizations (log$\xi$ $\sim$-1--0) than those that are outflowing ($\sim$200 km s$^{-1}$, log$\xi$ $\sim$1--2).
Future work identifying similar trends in other objects is crucial, since it and 
the potential inflows can suggest different interesting geometries, assuming comparable densities (see Figure \ref{fig:geos}).

One potential geometry (Option A in Figure \ref{fig:geos}) is that the `inflows' we report in 2004 (04-S2 and 04-S3) are not flowing towards the black hole in reality. Since we only measure a line-of-sight velocity, and the inclination of the system is $\sim$ 30--40$^{\circ}$ \citep{2025ApJ...995..200B, 2026ApJ..1003..103W},
it is plausible that these slow absorbers represent large gas clouds orbiting the black hole at a great distance, perhaps near or beyond the torus or the narrow line region. To explain the range in ionizations for these gases in PIE, the lower ionization gases may need to be in wider out-spiraling orbits, farther from the X-ray source.

Another potential geometry (Option B in Figure \ref{fig:geos}) is that the different components are stratified by ionization parameter, with the lower ionization components at greater distances, either not reaching the escape velocity of the black hole, and/or are slowed down by the ISM of the galaxy. This may lead some components to fall back towards the black hole.

Due to the similar velocity and line width of the $\sim$200 km s$^{-1}$ CIE absorber (04-S1) with the PIE absorber (04-S4), they may be co-spatially related, and the CIE absorber is shielded and/or shock-heated by the PIE absorber. 
It is also possible these gases have no relationship to each other and only have coincidentally similar velocities, line widths, and column densities.

In another interpretation, since our analysis best parses gases with discrete properties in equilibrium, it's also possible there is a less physically distinct collection of slow-moving non-equilibrium gases with a wider range of ionization parameters, like \citet{2010ApJ...708..981H} suggests. 

The ultra-fast, highly ionized outflows are potentially nearest to the central engine, and the cool-UFOs may have different origins, or be extensions of the hot UFO at later stages, but before they slow down to become warm absorbers (e.g. \citealt{2019A&A...627A.121S}).

\subsection{Outflow power}

To estimate the total wind power in each observation, we sum over the kinetic luminosity of each detected absorber, using the equation
\begin{equation} \label{KE}
    L_{KE} = \frac{1}{2} \dot{M}v^{2}
\end{equation} 
where $v$ is the velocity of the outflow and $\dot{M}$ is the mass outflow rate. Assuming a spherical outflow that is volume-filling to derive upper limits, we have
\begin{equation} \label{KE}
    \dot{M} = 4 \pi r N_H \mu m_{p} C_f v,
\end{equation} 

where $r$ is wind distance from the central source, $N_H$ is hydrogen column density, $\mu$ is mean molecular weight 1.4 for solar abundances, 
$m_p$ is the mass of a proton, and $C_g$ is the global covering fraction of the wind over the source (\citealt{2003ARA&A..41..117C},\citealt{2012ApJ...753...75C}). The kinetic luminosity is therefore 
\begin{equation} \label{KE}
    L_{KE} = 2 \pi r N_H \mu m_{p} C_f v^3.
\end{equation} 

We directly measure column density $N_H$ and the line-of-sight wind velocity $v$. To approximate the total fractional coverage of the wind over the X-ray source, with a spherical shape, we use the occurrence rate of ionized winds seen in population samples of AGN at random viewing angles, which is around $C_g \approx$ 0.5 \citep{1997MNRAS.291..403R, 1999ASPC..175..341C, 2014MNRAS.441.2613L}. 

To estimate the distance $r$ for photoionized absorbers, we use the geometric argument in which we assume the thickness of the absorber cannot exceed its distance from the central source \citep{1995MNRAS.273.1167R}. This means we can substitute $r$ $\geq$ $\Delta r$ in the definition of ionization parameter $\xi$,  
\begin{equation}\label{xi}
    \xi = \frac{L_{ion}}{r^{2}n_{H}},
\end{equation}
where $L_{ion}$ is the ionizing luminosity and $n_H$ is the hydrogen density. Then we can substitute $N = n \Delta r$ and derive an upper limit on the radial distance of a photoionized wind, or
\begin{equation} \label{r}
    r \leq \frac{L_{ion}}{N \xi}.
\end{equation}
If this produces a maximum radius larger than the galaxy ($>$ 10 kpc), we instead assume the wind is inside the galaxy. Using upper limits on radius, we can estimate the upper limit on the total photoionized wind power.

\begin{table}
	\centering
	\caption{Upper limit of maximum total kinetic outflow power of photoionized absorbers, as a percentage of $L_{Bol}$. The Ultra-fast outflow UF1  and all other absorbers potentially in CIE are considered in the scenario when  they are photoionized, since  their ionization  processes are ambiguous  or uncertain.}

	\label{tab:gtis}
	\begin{tabular}{lcc} 
		\hline
		Component(s) & 2000 & 2004 \\
		\hline
		Slow  (S1-S5) if PIE &  0.02$^{+0.04}_{-0.02}$ \% &  0.1$^{+0.05}_{-0.05}$ \%  \\
         Fast (F1) if PIE &  0.1$^{+0.01}_{-0.01}$ \% &  0.04$^{+0.01}_{-0.01}$ \%   \\
        Hot UFO (UF1) if PIE & $<$8 \% &   4$^{+11}_{-2}$ \% \\
		 \hline
		  total photoionized & $<$8 \% & 4 $^{+11}_{-2}$ \%\\
		\hline
	\end{tabular}
\end{table}

\subsubsection{Photoionized Outflow Power}
The maximum total kinetic outflow power of the winds in the 2004 dataset is 
$L_{KE}$ =  3.19$_{-0.03}^{+0.1}$ $\times 10^{43}$ erg $s^{-1}$, assuming they are all photoionized, and excluding cool-UFO absorbers.  We assume an ionizing luminosity of 
$L_{ion}$ = 2.4 $\times 10^{43}$ erg s$^{-1}$ (13.6 eV -- 13.6 keV, calculated from the SED in Section \ref{sec:sed}). This maximum photoionized wind power is 
4 $^{+11}_{-2}$\% of $L_{Bol}$,
assuming 
$L_{Bol}$ = 7.4 $\times 10^{44}$ (also from the SED in Section \ref{sec:sed}), and preserving the statistical uncertainty determined from our MCMC chains. This estimate is dominated $>$ 97\% by the power of the UFO (UF1) when we assume it is photoionized and launched at its maximum radius $\sim$0.0003 pc, near the inner disk. However, this component can be equally well modeled as collisionally ionized, potentially changing the result (see Section \ref{CIEpower}).
The estimated maximum photoionized total power is as much as $\sim$15\% $L_{Bol}$ within 1$\sigma$ confidence, which exceeds the minimum power simulations predict may be necessary to regulate bulge growth \citep[0.5-5\% $L_{bol}$,][]{2010MNRAS.401....7H}. Therefore, the photoionized X-ray winds of MCG-6 in 2004 could be significantly contributing to feedback within the galaxy.

The power of the detected photoionized absorbers in 2000 is negligible, $\sim$0.1\% $L_{Bol}$ or less due to the lack of any identified ultra-fast outflows. With an upper limit of $<$10$^{23.5}$ cm$^{-2}$, a similar hot photoionized UFO in 2000 could contribute as much as $<$8\% of the total bolometric power. This highlights that even with a nearby, bright object like MCG-6, a longer exposure time with HETG and/or observing with XRISM to constrain the winds in the hard X-ray band is crucial to constraining wind power. 

\subsubsection{Collisionally Ionized Outflow Power} \label{CIEpower}
The maximum power of the collisionally ionized absorber is not as easily constrained; since collisional level populations are independent of density, there is no simple limit argument we can make from the measured parameters. Furthermore, the occurrence rate of absorbers in CIE has not yet been studied, as they have only been considered in a few objects. This means the components in CIE not only have unknown radii, mass outflow rates, and covering fractions, but some may not necessarily be related to the AGN. It's possible that components in CIE, particularly with low velocities, should be excluded from AGN feedback discussions. This could yield large systematic uncertainties and inaccurate conclusions in total wind power if CIE absorption models are not considered.

\subsection{Importance of Completeness of Atomic Databases}\label{few}
Since an analysis like ours assumes that the atomic data for absorption models are complete and accurate, systematic uncertainties in atomic data can limit our results. Critically, missing atomic data can have large systematic effects on results, especially on components constrained by only a few atomic features in line-rich regions. For example, our results for the cool, relativistic (CUF1 and CUF2) components are constrained by the fewest features in the line-rich region of the soft X-ray band, and there is significant evidence that these are misidentified lines.

The same absorption lines were also identified as significant in \citet[][Holc10 hereafter]{2010ApJ...708..981H} in an ion-by-ion or line-by-line fitting methodology, but were interpreted as lines that don't require a relativistic blueshift. Using higher-order transition oxygen lines of the same ions, Holc10 instead attributes these features to warm absorbers. Their approach differs from our methodology because our self-consistent models utilize not only line shifts, as in line-by-line fitting, but also seek consistent line widths and column densities, thereby maximizing the constraining potential of the data. However, the databases underlying both analyses are different: \texttt{CLOUDY}, v. 23.01 uses Stout \citep{2015ApJ...807..118L} and Chianti for its atomic databases \citep{2023RMxAA..59..327C}, with UTA lines taken from openADAS, and these databases and/or our Cloudy-calculated models are likely missing the necessary higher order oxygen lines. Holc10 uses transition wavelengths adjusted from theoretical calculations using the Hebrew University Lawrence Livermore Atomic Code \citep[][HULLAC]{2001JQSRT..71..169B}, and many of which are now supported by laboratory measurements (particularly OV weaker resonances, \citealt[][]{2017MNRAS.465.4690M}), favoring their warm absorber interpretation over our relativistic one. 

More specifically, for the one broad feature at 19.5 \r{A}, our work finds an O III K$\alpha$ transition (rest frame 23.1 \r{A}), blueshifted by $\sim$45,000 km s$^{-1}$ (CUF2), while Holc10 attributes this line to a mix of an O V K$\gamma$ and O VI K$\beta$ transitions as part of the slow WAs.
Similarly, for the three features observed at 20.5, 20.1, 19.8 \r{A}, our framework fits them as O IV, O V, O VI K$\alpha$ transitions (rest frame 22.7, 22.3, 22.0 \r{A}) shifted by $\sim$30,000 km s$^{-1}$ (CUF1), while Holc10 finds these transitions to be part of the slow WAs and instead modeled by an unspecified O IV and an O V k$\beta$ transition.

\begin{figure*}
    \centering
	\includegraphics[width=\textwidth]{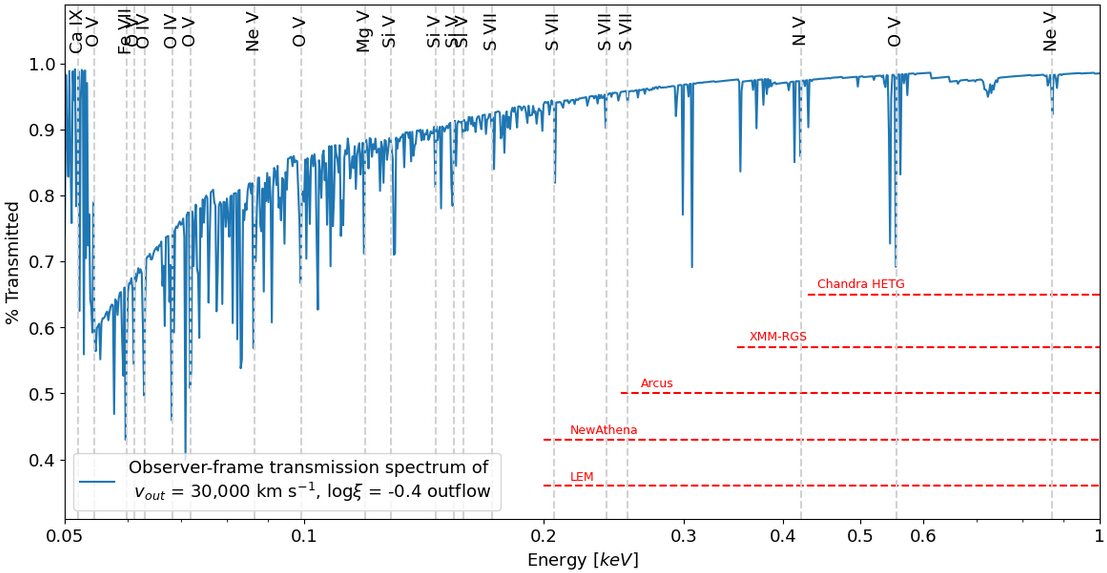}
    \caption{Transmission spectrum of cool UFO (CUF1) extended to energies in softer X-ray and UV, lower than HETG's range. Observations in this band from sufficiently long XMM-Newton observations or future X-ray missions like NewAthena, Arcus and LEM, could confirm these detections.}
    \label{fig:CUFs}
\end{figure*}

Due to the combination of relativistic speeds and low temperatures and/or ionization parameters, these absorbers, if they were physical, would not fit in the typical picture of WAs and UFOs, and would be most similar to the low-IP UFO type \citep{2024ApJS..274....8Y} seen in some AGN (e.g. \citealt{2013ApJ...772...66G, 2015ApJ...813L..39L, 2019A&A...627A.121S}). The prospect of low-IP UFOs is important to consider, since their existence could provide evidence of successful energy transfer or later evolutionary stage of hot UFOs at larger spatial scales. However, due to the systematic uncertainty to critical absorption lines in our models, the our relativistic interpretation is likely incorrect, but nor can we robustly reproduce a non-detection. 

Precise modeling of the K$\beta$, K$\gamma$ O IV-VI K-shell transition energies and cross sections is necessary to robustly access potential low-IP UFOs in X-ray gratings and microcalorimeters in the future. Another approach is to constrain them using more lines, which are largely in softer X-ray and/or UV range. For example, the fitted components from this work absorb primarily in the UV or soft X-rays ($<$0.5 keV), a small part of which is observed by HETG. In Figure \ref{fig:CUFs} we show the transmission spectrum for an absorber resembling the 30,000 km~s$^{-1}$ cool UFO (CUF1), which highlights that the majority of the absorption of such a component would occur at UV at energies. Obtaining sufficient UV constraints is likely not possible in MCG-6 due to extreme reddening \citep{1997MNRAS.291..403R}. XMM-Newton's RGS can observe energies lower than HETG, which, with long exposures, may be sufficient to constrain N V lines at $\sim$0.4 keV for tentative cool-UFOs with similar properties (see Figure \ref{fig:CUFs}). Alternatively, future X-ray instruments that can observe in even softer X-rays, such as proposed missions like NewAthena \citep{2025NatAs...9...36C}, Arcus \citep{2023HEAD...2030604B}, and LEM \citep{10.1117/12.3020570}, could robustly determine the validity of relativistic low ionization winds.

\subsection{Other Limitations}
In this section, we discuss how our assumptions may impact the results of this work, and which may be useful for further study, but are beyond the scope of this present work.

\subsubsection{Time-averaged Spectral Modeling}
AGN are known to vary in luminosity on nearly every timescale: minutes, hours, days, weeks, years, or more. MCG-6 displays no extreme flare events over the course of each observation \citep{2002ApJ...570L..47L, 2005ApJ...631..733Y}. However, works such as \citet{2026ApJ..1003..103W} have shown that some relativistic reflection parameters  (e.g., reflection fraction, coronal height/emissivity profile, photon index, disk ionization) may vary on timescales shorter than an observation. There is also evidence that many of the fastest outflows ($>$ 1000 km s$^{-1}$) in a given AGN, including MCG-6, may vary on timescales of days or kiloseconds \citep[][]{2020MNRAS.493.1088I, 2022ApJ...940..122R}. Our work focuses on variability on timescales of $\sim$months or larger and leveraging spectra averaged over shorter timescales to gather sufficient signal-to-noise.

\subsubsection{Solar Abundance Assumption}
Our work assumes the metallicity of absorbing zones to be solar. This means that if the abundances are lower or higher, the inferred column densities would change. Similarly, if the elemental abundance ratios are non-solar, the inferred equilibrium temperature or ionization parameter may be biased.

In particular, for the 00-F1/04-F1 absorber at $\sim$1900 km $s^{-1}$, the relative abundance of Fe XXV and Fe XXVI to lower z metals and their lower energy ionic transitions may be significant to the preference of the fit (see Figure \ref{fig:fe2425}).

\subsubsection{Equilibrium Assumption}
To use models of photoionization equilibrium, we assume the recombination timescale of a gas is short relative to the variability of the primary ionizing continuum. At low densities, 
the recombination timescale may become relevant, a possibility we do not explore in this work, nor do we obtain constraints on 3-dimensional density, leaving uncertainty on the assumption of pure equilibrium. Time-dependent photoionization modeling may lead to differences in measured parameters (\citealt{2023ApJ...946...93S}, \citealt{1999ApJ...512..184N}, \citealt{2022ApJ...940..122R}).

\subsubsection{Systematic uncertainty from the SED}
Due to extensive reddening due to dust in MCG-6 \citep{1997MNRAS.291..403R}, the UV portion of the spectrum is highly uncertain. We assume a typical $\alpha_{ox} = 1.5$ (e.g., \cite{2018A&A...619A..95C}) to estimate the UV flux; however, $\alpha_{ox} = 1.2-1.7$ are also common within 1--2$\sigma$. 

A different UV flux in the SED will affect the measured parameters for the photoionized models, especially the ionization parameter, by less than an order of magnitude for a given absorber. For the fastest, hot absorber, we test a lower luminosity SED with an $\alpha_{ox}$ = 1.25, and find an ionization parameter different by $\sim$10\% for the PIE case.

\subsubsection{Systematic uncertainty from the atomic data} \label{atm}
Our framework assumes that the atomic data for the neutral and ionized models are complete and accurate. The three major photoionization codes: Cloudy, XSTAR \citep{2024EPJD...78...93M}, and PION \citep{2021ascl.soft03024M}, use different atomic databases, and \citet{2016A&A...596A..65M} has shown that the difference in inferred parameters is typically 10-20\%. However, missing or inaccurate atomic data can still have systematic effects on our results, especially on components constrained by only a few atomic features. For our results constrained by the least features in the line-rich region of the soft band, the CUFs, we discuss alternative results in Section \ref{few}.

\section{Summary}\label{conc}

We present a time-averaged Bayesian analysis of the archival \emph{Chandra} HETG data of MCG-6-30-15 (MCG-6) to study its ionized outflows.
After modeling the broadband emission, we use \texttt{Cloudy} to calculate large numerical models of absorbing gas with equal atomic levels in both collisionally ionized equilibrium (CIE) and photoionized equilibrium (PIE), such that the sampling of each parameter (ionization parameter or temperature, column density, and line width) matches the resolution of the data. We use the Deviance Information Criterion (DIC) to determine whether an absorber in PIE or CIE better represents the data for each potential component, and cease adding components if none can yield a statistically significant DIC improvement, without over-fitting.

In total, we detect 11 absorbing zones in the 2004 observation and 6 absorbing zones in the shorter 2000 observation. These include:
\begin{itemize}
\item \textbf{
First Detections of Potential, Slow CIE Absorbers.
}
We find the second most statistically significant absorbing component has a velocity near the rest frame of MCG-6, that can be equally well modeled as a log($\xi$/[erg cm s$^{-1}$] $\sim$2 component in PIE, likely as a part of the wide range of warm absorbers, or as a log($T$/[K]) $\sim$ 6 component in CIE, a state often not considered in similar analyses. This component is stable across four years and its temperature in the CIE case is consistent with galactic and/or CGM absorption of a MCG-6-mass galaxy, potentially indicating this absorber is not associated with the AGN.
Our results add to the increasing evidence that outflows in CIE may be a common \citep[see also][]{2022MNRAS.516.5027O, 2025ApJ...979..101T, 2025arXiv250111562M} but often unidentified wind type, affecting studies of outflow demographics and AGN feedback.

\item \textbf{Warm Absorbers with a Possible Velocity and Ionization Gradient.} 
 Assuming all winds are in PIE, find a consistent system of warm absorbers that vary from low-ionization and inflowing velocities (log($\xi$/[erg cm s$^{-1}$]) $\sim -$0.5, $v_{in} \sim -$300 km s$^{-1}$) linearly to hotter mild outflows (log($\xi$/[erg cm s$^{-1}$]) $\sim -$2, $v_{out} \sim$  300 km s$^{-1}$). We interpret these either as rotating in a wide out-spiraling orbit or turning around at greater distances and flowing back towards the black hole.

\item \textbf{New Nature and Variability of Well-Known 1900 km s$^{-1}$ Component.}
We find that the moderately fast $\sim$1900 km s$^{-1}$ outflow, seen previously in MCG-6,  may be better represented by CIE given the relative ion fractions, and may increase in temperature and column density over four years. 

\item \textbf{First Detection of Hot UFO in  MCG-6-30-15 Grating Data.} 
In the deeper 2004 dataset, we independently detect a wind moving at $\sim$ 0.08c, only otherwise seen in recent XRISM work and in XMM-Newton CCD variability analysis. We find that both CIE and PIE can equally well explain this component. If this UFO is in PIE and exhibits its maximum possible power, it may play a significant role in AGN feedback, with the power $L_{KE}\sim 5\%L_{Bol}$. This detection highlights the relevance of HETG for UFO studies, especially if paired with sensitive statistical approaches.

\item \textbf{Dusty Warm Absorbers in MCG-6.} Similar to previous works, we find the significant spectral features around $\sim$0.7 keV are well represented as cold Fe L-shell features, likely that of neutral and/or dusty Fe, in the rest frame of MCG-6. We measure an anomalous O:Fe ratio of $\sim$0.25, which strongly suggests that not all Fe is in the ISM but that large fractions of Fe are embedded in winds close to the AGN. To definitively resolve the nature of the cold Fe in MCG-6, precise measurements of the Fe edge in molecules and low ionization gases are needed. 

\item \textbf{Systematic Uncertainty in Atomic Data Limiting Results.} In the 2004 data, we find statistically significant soft band features, including 3 narrow and 1 mildly broadened line, unaccounted for by any slow absorbing zones with our calculated \texttt{Cloudy} models.
Although our framework finds them 
represented as low-ionization, 
cool relativistic components, there is significant theoretical and experimental evidence that these are instead non-relativistic features of atomic data missing from our warm absorber models. The prospect of cool relativistic absorbers is potentially important to connecting QSO winds, WAs, UFOs and how they transfer AGN feedback, however to robustly conclude detections or non-detections of such components, self-consistent ionization models require complete atomic datasets.

\end{itemize}

Our framework demonstrates that deep archival Chandra HETG data can be used to study hot UFOs and naturally provide a temporal baseline for XRISM and future measurements.
We also show that measurements of these deepest HETG datasets can be limited by the completeness of tested models, here for higher order ionized O K-shell transitions, Fe edges of molecules and low ionization/neutral gases. 
Our work highlights the importance of viewing winds in both the hard and soft X-rays, as well as going deeper with future missions like NewAthena \citep{2025NatAs...9...36C}, Arcus \citep{2023HEAD...2030604B}, and LEM \citep{10.1117/12.3020570} to test the connection between quasar/UV winds and local X-ray winds.

\section*{Acknowledgments}
EH thanks the reviewer for their helpful comments. 
EH especially thanks Liyi Gu for his help ensuring our photoionization models contained sufficient atomic levels.
EH extends special thanks to the CIAO and HEASOFT helpdesk's Kenny, Nick and Bryan for their technical assistance.
EH also thanks Javier Garcia for answering inquiries about the \emph{relxill} model family package, and Keith Arnold and Craig Gordon for their help with \texttt{xspec}.

EH and AO thank Joern Wilms for his insightful discussions.
AO thanks Natalie Hell for useful discussions.

Resources supporting this work were provided by the NASA High-End Computing (HEC) Program through the NASA Center for Climate Simulation (NCCS) at Goddard Space Flight Center.

This research has made use of software provided by the Chandra X-ray
Center (CXC) in the application package CIAO.
The scientific results reported in this article are based on data obtained from the Chandra Data Archive.

The material is based upon work supported by NASA under award number 80GSFC24M0006.  Support for this work was provided by the National Aeronautics and Space Administration through Chandra Award Number TM1-22006X issued by the Chandra X-ray Observatory Center, which is operated by the Smithsonian Astrophysical Observatory for and on behalf of the National Aeronautics Space Administration under contract NAS8-03060. 


\appendix

\section{\texttt{Cloudy} Model Specifics}\label{clo}
 We assume all components not to be in a high-density regime, such that the density may affect the ionization balance, by specifying a hydrogen density $N_{H} = 10^5$ cm$^{-3}$. We use the default solar abundances of \texttt{Cloudy}, which use a combination of \citet{2001ApJ...556L..63A, 2002ApJ...573L.137A}, \citet{2001AIPC..598...23H}, and \citet{1998SSRv...85..161G}, outlined in-depth in the c25.00 manual Hazy 1, Section 7.1. We also assume the default plane parallel, open geometry, and iterate the code until the desired column density is reached.

We increase the internal spectral resolution of the code to be high in the relevant energy ranges of HETG. The default, which we keep everywhere else in the spectrum, is a resolving power of R = 300, which we change to R = 3000 for 11-1100 Rydbergs ($\sim$0.15-15 keV)

\texttt{CLOUDY}, v. 25.00 uses Stout \citep{2015ApJ...807..118L} and Chianti for its core atomic databases \citep{2023RMxAA..59..327C}, with UTA lines taken from openADAS.


\bibliography{sample701}{}
\bibliographystyle{aasjournalv7}

\end{document}